\documentclass[preprint]{revtex4}
\usepackage{enumerate}
\usepackage{graphicx}
\usepackage{dcolumn}
\usepackage{bm}
\usepackage{amsmath}
\usepackage{amsxtra}
\usepackage{amstext}
\usepackage{amssymb}
\usepackage{latexsym}

\pdfoutput=1
\begin{document}

\title{Derivation of Reference Distribution Functions for Tokamak-plasmas by Statistical Thermodynamics}

\author{
Giorgio SONNINO$^{1*}$, Alessandro CARDINALI$^{2}$, Philippe PEETERS$^{1}$, Gy\"{o}rgy STEINBRECHER$^{3}$, Alberto SONNINO$^{4}$
\vskip0.1truecm
$^{1}$Universit{\'e} Libre de Bruxelles (U.L.B.), Department of Theoretical Physics and Mathematics, Campus de la Plaine C.P. 231 - Bvd du Triomphe, 1050 Brussels - Belgium\\
$^{2}$EURATOM-ENEA Fusion Association, Via E.Fermi 45, C.P. 65 - 00044 Frascati (Rome) - Italy\\
\vskip0.1truecm
$^{3}$EURATOM-MEdC Fusion Association, University of Craiova, Faculty of Exact Sciences, Str.A.I.Cuza Street 13, Craiova-200585, Romania\\
\vskip0.1truecm
$^{4}$ Universit{\'e} Catholique de Louvain (UCL), Ecole Polytechnique de Louvain (EPL), Rue Archim$\grave{\rm e}$de, 1 bte L6.11.01, 1348 Louvain-la-Neuve, Belgium}

\begin{abstract}

A general approach for deriving the expression of reference distribution functions by statistical thermodynamics is illustrated, and applied to the case of a magnetically confined plasma. The local equilibrium is defined by imposing the minimum entropy production, which applies only to the linear regime near a stationary thermodynamically non-equilibrium state and the maximum entropy principle under the scale invariance restrictions. This procedure may be adopted for a system subject to an arbitrary number of thermodynamic forces, however, for concreteness, we analyze, afterwords, a magnetically confined plasma subject to three thermodynamic forces, and three energy sources: i) the total Ohmic heat, supplied by the transformer coil, ii) the energy supplied by Neutral Beam Injection (NBI), and iii) the RF energy supplied by Ion Cyclotron Resonant Heating (ICRH) system which heats the minority population. In this limit case, we show that the derived expression of the distribution function is more general than that one, which is currently used for fitting the numerical steady-state solutions obtained by simulating the plasma by gyro-kinetic codes. An application to a simple model of fully ionized plasmas submitted to an external source is discussed. Through kinetic theory, we fixed the values of the free parameters linking them with the external power supplies. The singularity at low energy in the proposed distribution function is related to the intermittency in the turbulent plasma.

\vskip0.7truecm
\noindent *Email: gsonnino@ulb.ac.be
\vskip0.5truecm 
\noindent 

\end{abstract}

\maketitle








\section{Introduction}\label{intr}
\vskip0.2truecm 
\noindent 
\noindent Statistical thermodynamics constitutes a powerful tool for deriving the reference density distribution functions (DDF), ${\mathcal{F}}^{R}$ (see, for instance, \cite{balescu0}). By definition, the reference DDF is an initial DDF, referred to as ${\mathcal F}^R$, which should depend only on the invariants of motion, with the property to evolve slowly from the local equilibrium state (LES) i.e., the reference DDF remains confined for a sufficiently long time. Hence, ${\mathcal F}^R$ results in a perturbation of the local equilibrium state. We shall use Prigogine's statistical thermodynamics to derive ${\mathcal{F}}^{R}$ for open plasmas systems close to a local equilibrium state. The LES is defined by assuming the validity of a minimal number of hypotheses: the minimum entropy production principle (MEP) and the maximum entropy principle (MaxEnt principle) under two scale invariance restrictions. We recall that the MEP establishes that, in the Onsager region, if the matrix of the transport coefficients is symmetric, a thermodynamic system relaxes towards a stable steady-state in such a way that the rate of the {\it entropy production strength}, $\sigma$, is negative. The inequality is saturated at the steady-state. As mentioned above, the MaxEnt principle is submitted to constraints. In this regard, it is important to stress that the set of restrictions, imposed by very general physical principles, including scale invariance, gives automatically, for some range of parameters, a distribution function whose singularity can be interpreted in the terms of intermittency on turbulent plasma \cite{sonninoPRE1}.

\noindent The density probability distribution of finding a state in which the values of the fluctuating thermodynamic variable, $\beta_\kappa$, lies between $\beta_\kappa$ and  $\beta_\kappa+d\beta_\kappa$ is
\begin{equation}\label{i1}
{\mathcal F}={\mathcal N}_0\exp[-\Delta_I S]
\end{equation}
\noindent where ${\mathcal N}_0$ ensures normalization to unity, and we have introduced the dimensionless entropy production $\Delta_I S$ \cite{prigogine}. Suffix $I$ stands for {\it irreversibility}. By introducing the entropy production variation due to fluctuations, Prigogine proposes Eq.~(\ref{i1}), which is valid for open thermodynamic systems. This equation generalizes the Einstein theory of fluctuations which, on the contrary, applies only to adiabatic or isothermal transformations. Note that ${\mathcal F}^R$ that we want to determine, is a particular case of non-equilibrium density probability distributions, hence it can also be brought into the form (\ref{i1}). The negative sign in Eq.~(\ref{i1}) is due to the fact that, during the processes, $-\Delta_IS\leq0$ \cite{prigogine}. Indeed, if $-\Delta_IS$ were positive, the transformation $\beta_\kappa\rightarrow \beta'_\kappa$ would be a spontaneous irreversible change and thus be incompatible with the assumption that the initial state is a {\it stable} (local) equilibrium state \cite{prigogine}, \cite{prigogine1}. We suppose that the system is subject to ${\tilde N}$ thermodynamic forces. The thermodynamic forces defined as $X^\kappa={\partial \Delta_IS}/{\partial \beta_\kappa}$, and the thermodynamic flows defined as $J_\kappa={d\beta_\kappa}/{dt}$, are linked each others by the following equations \cite{degroot}
\begin{equation}\label{i2}
\frac{d_IS}{dt}=\sum_{\kappa=1}^{\tilde N}  X^\kappa J_k\!\geq 0
\end{equation}
\noindent Notice that $d_IS$ is {\it not} an exact differential. We also recall that
\begin{equation}\label{i2a}
\Delta_I S=\int^\beta_{\beta^{eq.}}d_IS\qquad{\rm with}\qquad \frac{d_IS}{dt}=\int\sigma d{\bf x}
\end{equation}
\noindent with $d{\bf x}$ denoting the {\it spatial volume element} and the integration is over the whole volume occupied by the system. Note that the probability density function (\ref{i1}) remains unaltered for
flux-force transformations, $X_{\kappa }\rightarrow X_{\kappa }^{\prime }$
and $J_{\kappa }\rightarrow J_{\kappa }^{\prime }$, leaving invariant the
entropy production \cite{sonninoPRE} 
\begin{equation}\label{i3}
\frac{d_{I}S}{dt}=\sum_{\kappa =1}^{\tilde N}J_{\kappa }X^{\kappa }=\sum_{\kappa
=1}^{\tilde N}J_{\kappa }^{\prime }X^{\prime }{^{\kappa }}
\end{equation}
\noindent As stated above, the explicit determination of ${\mathcal{F}}^{R}$ requires the preliminary knowledge of the local equilibrium state (being ${\mathcal{F}}^{R}$ a perturbation of LES). LES should be defined uniquely by imposing a set of minimum conditions. These conditions should be established according to the particular physical situation that we are analyzing. It is important to stress that the set of restrictions, imposed by very general physical principles, including scale invariance, gives automatically, for some range of parameters, a distribution function whose singularity can be interpreted in the terms of intermittency on turbulent plasma (see Appendix). This procedure is quite general and it can be applied to a system
subject to an arbitrary number of thermodynamic forces. However for
concreteness, we limit ourselves to study the particular case of a system
submitted to only three thermodynamic forces and to three energy sources : i) Ohmic heat, ii) Neutral Beam Injection (NBI), and iii) RF ICRH power supplied.

\noindent In this paper, we shall analyze plasmas subject to three thermodynamic forces by stressing that this work is not restricted to, but applied to, tokamak plasmas. Tokamak-plasmas are therefore studied as a concrete example of calculation. 

\noindent We mention that the present state of the art in tokamak modeling allows
obtaining distribution functions from sophisticated numerical codes (e.g.,
 TORIC+SSFPQL for ICRH external injection \cite{bilato}). These numerical
codes can actually be used for calculating numerically the particle
distribution function for different power injection schemes and different
levels of approximations in treating sources, collision operators and
particle motion in the equilibrium fields. The results of the present work also provide a reasonably complete class of model distribution functions to be
used in gyrokinetic or hybrid gyrokinetic-fluid simulations of interest for magnetically confined plasmas. The
coupling of such gyrokinetic or hybrid gyrokinetic-fluid simulation with
power deposition codes, like those mentioned above, is a challenging task,
which is underway. In the meantime, sufficiently accurate description of
model particle distribution functions are needed, which nowadays are
typically chosen as "reasonable" model functions of the particle constants
of motion (often just a Maxwellian, as in GTC and GYRO, or more generally an
anisotropic Maxwellian or slowing down for the case of Hybrid Magnetohydrodynamic Gyro-kinetic Code (HMGC), M3D and NOVA) \cite{cardinali1}-\cite{zonca}.
In this work, the constraints that non-equilibrium statistical mechanics
impose on the adopted model distribution functions are derived and a class
of model distribution function is proposed, whose usefulness is therefore,
the readiness to be adopted in gyrokinetic or hybrid gyrokinetic-fluid
simulations of fusion interest. However, at the present time, none of the above-mentioned codes can
be used in the same way, for the intrinsic difficulty of an actual
integrated simulation. The advantage of the model distribution functions
obtained in this work is therefore evident. The fundamental issue here
is that the detailed application to tokamaks comes in only when the specific
form of constants of motion in the equilibrium fields is adopted. Before
that point, there is a construction of the particle distribution function
out of the equilibrium, as we expect that to be in the case of a tokamak. Indeed, starting from  
an arbitrary initial state, collisions would tend, if they were alone, to bring the system very quickly to a stationary state. But the slow processes i.e., the free flow and the electromagnetic processes, prevent the plasma from reaching this state. The result is that, after a short time, the plasma reaches a state {\it close} to the {\it local equilibrium}. This state is referred to as the {\it reference state}. From here on, the distribution function evolves on the slow time scale. Notice that the local equilibrium state is {\it not} a state of thermodynamic equilibrium, because the latter must be homogeneous and stationary.

\noindent The paper is organized as follows. In Section~(\ref{ThermDF}) we
derive the general expression of the reference density of distribution
probability, ${\mathcal{F}}^{R}$, by a pure thermodynamic approach. The
parameters, entering in the expression of ${\mathcal{F}}^{R}$, are
determined by kinetic theory in Section~\ref{parameter} by adopting a model
for the tokamak-plasma and the external sources. Section~\ref{estimation} addresses the following questions :

\begin{description}
\item ${\bullet}$ \textit{For collisional tokamak-plasmas, how much is the
deviation of the reference DDF, ${\mathcal{F}}^R$, from the Maxwellian} ?

\item ${\bullet}$ \textit{Does this deviation coincide with the one
estimated by the neoclassical theory} (see, for example, Ref.~\cite{balescu2}
) ?
\end{description}

\noindent In other words, we should ensure that the expression that we found for the
reference DDF, ${\mathcal{F}}^R$, coincides exactly with the one
predicted by the neoclassical theory for collisional tokamak-plasmas. We shall see that the answers are affirmative and, at the same time, such an identification will allow determining the free parameters appearing in the reference DDF. Some concluding remarks
can be found in the Section~\ref{cs}. 
\vskip 0.2truecm

\vskip 0.2truecm

\section{Thermodynamic Derivation of the Distribution Function. General Considerations.}\label{ThermDF} 
\vskip0.2truecm 
\noindent In this section, we derive the form of ${\mathcal{F}}^{R}$ by following a purely thermodynamic approach. As usual, the gyro-kinetic (GK) theory makes often use of an initial distribution function of guiding centers. In the GK simulations, as well as in the GK theory, this initial distribution function is usually taken as a reference DDF if it depends only
on the invariants of motion and it evolves slowly from the local equilibrium
state i.e., in such a way that the guiding centers remain confined for
sufficiently long time. After a short transition time, the state of the
plasma remains close to the reference state, ${\mathcal{F}}^{R}$, which results in a small deviation of the local equilibrium state (LES). The expression of the
coefficients of the ${\mathcal{F}}^{R}$ will be determined in the next
section by kinetic theory. The reference DDF is obtained by perturbing the
local equilibrium state. The procedure reported in Ref.~\cite{sonninoPRE1} refers to an open system subject to ${\tilde N}$ thermodynamic forces with the local equilibrium state determined by the following two conditions.
\noindent 
\begin{description}
\item {\it {\bf i)} The local equilibrium state corresponds to the values of the $N$ Prigogine$'$s type (fluctuating) variables $\beta_i$ (with $N< {\tilde N}$) for which the entropy production tends to reach an extreme}.
\end{description}
\noindent This special class of variables $\beta_i$ will be denoted as $\alpha_i$. Hence, $\alpha_i$ with $i=1,\cdots, N< {\tilde N}$, are the fluctuating variables $\beta_i$ of Prigogine's type. By definition, a fluctuation is of Prigogine 's type if the entropy production is expressed in quadratic form with respect to these fluctuations (for an exact definition of PrigogineÕs fluctuations refer to Refs~\cite{prigogine}, \cite{degroot}). Under this assumption, close to the local equilibrium and around the extreme value $\partial\Delta_IS/\partial{\alpha_\kappa}\mid_{\alpha_1\cdots \alpha_N=0}\ =0$ (with $\kappa=1,\cdots ,N$), the entropy production can be brought into the form
\begin{equation}\label{i4}
-\Delta_IS=g_0(\bar\beta)-\frac{1}{2}\sum_{i,j=1}^Ng_{ij}(\bar\beta)\alpha_i\alpha_j+h.o.t.
\end{equation}
\noindent Here, ${\bar\beta}$ stands for the vector $(\beta_{N+1},\cdots, \beta_{\tilde N})$ and $h.o.t.$ for {\it higher order terms}. Hence, ${\bar\beta}$ denotes the set of fluctuations, which are {\it not} of Prigogine's type. Notice that the general DDF, ${\mathcal F}$, becomes a reference DDF ${\mathcal F}^{R}$ when the expression of entropy production is given by Eq.~(\ref{i4}). The DDF related to the variables $\bar\beta$, at $\alpha_i=0$ (with $i=1\cdots N$), reads
\begin{equation}\label{i5}
\mathcal{P}(\bar\beta)\equiv{\mathcal F}^{R}\mid_{\alpha_1\cdots \alpha_N=0}={\mathcal N}_0\exp[-\Delta_IS\mid_{\alpha_1\cdots \alpha_N=0}]={\mathcal N}_0\exp[g_0(\bar\beta)]
\end{equation}
\noindent $\mathcal{P}(\bar\beta)$ is determined by the following condition.
\begin{description}
\item {\it {\bf ii)} At the extremizing values $\alpha_i=0$ with $i=1,\cdots N$, under the scale invariance restrictions, the system tends to evolve towards the maximal entropy configurations}.
\end{description}
\noindent Coefficients $g_{ij}$ are directly linked to the transport coefficients of the system \cite{sonninoPRE1}. With these coefficients we may form a positive definite matrix, which can be diagonalized by obtaining 
\begin{equation}
-\Delta _{I}S=g_{0}(\bar\beta)-\sum_{i,j=1}^{N}\delta _{ij}c_{i}(\bar\beta)(\zeta _{i}-\zeta _{i}^{0})^{2}+h.o.t.  \label{i6}
\end{equation}
\noindent where $\delta _{ij}c_{i}({\bar\beta})$ is a positive definite matrix and $\delta _{ij}$ denotes Kronecker$'$s delta. Eq.~(\ref{i6}) allows describing the entire process in terms of $N$ \textit{independent} processes linked to the $N$ independent fluctuations $\zeta _{1},\cdots, \zeta_{N}$. The expression of the reference density of
distribution function is now expressed through a set of convenient variables $\{\zeta _{i}\}$ (with $i=1,\cdots , N$) of the type, \textit{degrees of advancement} (for a rigorous definition of these variables see, for example, Ref.~\cite{prigogine1}. See also the footnote 
\footnote{We recall that, by definition, the degrees of advancement variables $\zeta
_{j}$ satisfy the condition $\lim_{\zeta _{j}\rightarrow \zeta _{j}^{0}}\xi
_{i}=0$ \cite{prigogine}.})

\noindent In this paper, we shall restrict ourselves by analyzing plasmas subject to three thermodynamic forces. Magnetically confined plasmas are therefore studied as a concrete example of calculation. In the case of an axisymmetric magnetically confined plasma, after having performed the guiding center transformation, the necessary variables for describing the system reduce to four independent variables \cite{balescu1}. These variables are defined as follows. One of these ones is the {\it poloidal magnetic flux}, $\psi$, which for simplicity we consider not to be a fluctuating variable. Ultimately, plasma is subject to three thermodynamic forces (i.e., ${\tilde N}=3$), linked to the three (fluctuating) variables. One of these latter variables is the {\it particle kinetic energy per unit mass}, $w$, defined as $w=(v_{\parallel}^{2}+v_{\perp }^{2})/2$ with $v_{\parallel }$ denoting the parallel component of particle's velocity (which may actually be parallel or antiparallel to the magnetic field), and $v_{\perp }$ the absolute value of the perpendicular velocity \cite{balescu1}. The remaining two fluctuating variables are the \textit{toroidal angular moment}, $P_{\phi }$, the variable, $\lambda $. These quantities are defined as (for a rigorous definition, see any standard textbook such as, for example, \cite{balescu1}) 
\begin{align}\label{ddf1}
&P_{\phi }=\psi +\frac{B_{0}}{\Omega _{0c}}\frac{Fv_{\parallel }}{\mid B\mid}\equiv\zeta_1\\
&\lambda\equiv\frac{\mu }{w}=\frac{\sin ^{2}\theta _{P}}{2\mid B\mid }\equiv\zeta_2\qquad 
\mathrm{with}\quad \mu =\frac{v_{\perp }^{2}}{2\mid B\mid }
\end{align}
\noindent Here $\Omega _{0c}$ is the \textit{cyclotron frequency} associated with the magnetic field along the magnetic axis, $B_0$. $\mid B\mid $, $F$ and  $\theta _{P}$ denote the magnetic field intensity, the {\it characteristic of axisymmetric toroidal field} depending on the surface function $\psi$ and the \textit{pitch angle}, respectively. $P_{\phi }$ and $\lambda$ are considered as two Prigogine$'$s variables. Notice that, even though these variables depend on $w$, actually their variations are independent with each other. So $P_\phi$, $\lambda$ and $w$ are three {\it independent} variables \cite{balescu1}. We define our LES according to the conditions {\bf i}) and {\bf ii}), reported in Section {\ref{intr}, submitted to the two-scale invariant restrictions $\mathrm{E}[w]= {\rm const.} >0$ and $\mathrm{E}[\ln (w)]={\rm const.}$ (where $\mathrm{E}[\ ]$ is the expectation operation). Under these conditions, the DDF for the $w$ variable is given by a {\it gamma distribution function} \cite{park}-\cite{sonninopreb}
\begin{equation} \label{ddf3}
{\mathcal{P}}(w/\Theta )={\mathcal{N}}_{0}\Bigl(\frac{w}{\Theta }\Bigr)
^{\gamma -1}\exp [-w/\Theta ]
\end{equation}
\noindent where we have introduced the \textit{scale parameter} $\Theta $ and the \textit{shape parameter} $\gamma $. The motivation for the choice of the two-scale invariant restrictions as well as the special mathematical properties of the resulting DDF can be found in Appendix (\ref{just}) and in Ref.~\cite{sonninoPRE1}. In the volume element $d{\widehat{\Gamma}}=d\psi dwdP_\phi d\lambda d\phi d\Phi$, where $\phi$ and $\Phi$ are the {\it toroidal angle} and the {\it gyro-phase angle}, respectively, the reference state takes the form $d{\mathcal{\widehat{F}}}^{R}={\mathcal{F}}^{R}d{\widehat{\Gamma}}$ with
\begin{equation}\label{ddf4}
d{\mathcal{\widehat{F}}}^{R}={\mathcal{N}}_{0}\Bigl(\frac{w}{\Theta }\Bigr)
^{\gamma -1}\!\!\!\!\!\!\exp [-w/\Theta ]\exp [-c_{1}(w/\Theta )(P_\phi-P_{\phi0})^2]\exp [-c_{2}(w/\Theta )(\lambda-\lambda_{0})^{2}]\!\mid {\mathcal{J}}\mid \ \!\!d{\widehat{\Gamma}}
\end{equation}
\noindent where the scripts $0$ refer to (local) equilibrium values. The phase space volume element $d\Gamma =d\mathbf{x}d\mathbf{v}$ is linked to $d{\widehat{\Gamma }}$ by 
\begin{equation}\label{ddf5}
d\Gamma =\mid \!{\mathcal{J}}\!\mid d{\widehat{\Gamma }} 
\end{equation}
\noindent with $\mid \!{\mathcal{J}}\!\mid $ denoting the Jacobian between $ d\Gamma $ and $d{\widehat{\Gamma }}$. If  we interpret our reference DDF as a time and ensemble average of the physical DDF describing turbulent plasma, then the singularity at $w=0$ for $0<\gamma<1$ can be related to the intermittency \cite{sonninoPRE1}. Notice that at the point with coordinates $(P_\phi,\lambda,w)=[P_{\phi 0},\lambda_0,(\gamma -1)\Theta ]$, the system satisfies the principle of maximum entropy and the entropy production reaches its extreme value. Let us now suppose that $c_{1,2}(w/\Theta)$ are narrow coefficients with small deviations from the expectation value. In this situation we may expand coefficients $c_{1}$ and $c_{2}$ up to the leading order in $w/\Theta$. By taking into account that $P_\phi^2\sim v_\parallel^2$ and $\lambda\sim v_\perp^2/w$, we get
\begin{align}\label{ddf6}
c_{1}(w/\Theta )& \simeq c_{1}^{(0)}\equiv \Bigl(\frac{1}{\Delta P_{\phi }}\Bigr)
^{2}=const. \nonumber\\
c_{2}(w/\Theta )& \simeq c_{2}^{(0)}+c_{2}^{(1)}\frac{w}{\Theta }\equiv\frac{1}{\Delta \lambda _{0}}\Bigl(\frac{\Delta \lambda
_{0}}{\Delta \lambda _{1}}+\frac{w}{\Theta }\Bigr)\geq 0
\end{align}
\noindent where $\Delta P_{\phi }$, $\Delta \lambda _{0}$ and $\Delta \lambda _{1}$ are constants. Finally, the expression for the density distribution function ${\mathcal{F}}^{R}$ reads 
\begin{equation}\label{ddf7}
{\mathcal{F}}^{R}={\mathcal{N}}_{0}\Bigl(\frac{w}{\Theta }\Bigr)^{\gamma
-1}\!\!\!\!\!\!\exp [-w/\Theta ]\exp \Bigl[-\Bigl(\frac{P_{\phi }-P_{\phi 0}
}{\Delta P_{\phi }}\Bigr)^{2}\Bigr]\exp \Bigl[-\Bigl(\frac{\Delta \lambda
_{0}}{\Delta \lambda _{1}}+\frac{w}{\Theta }\Bigr)\frac{(\lambda -\lambda
_{0})^{2}}{\Delta \lambda _{0}}\Bigr]\mid {\mathcal{J}}\mid 
\end{equation}
\noindent where ${\mathcal{N}}_{0}$ ensures normalization to unity 
\begin{equation}\label{ddf8}
\int_{\widehat{\Omega}}d{\mathcal{\widehat{F}}}^{R}=\int_{{\widehat{\Omega}}}{\mathcal{F}}
^{R}d{\widehat{\Gamma}}=1  
\end{equation}
\noindent with $\widehat\Omega$ denoting the phase space-volume in the $\widehat\Gamma$ space. The presence of the free parameter $c_{2}^{(0)}$ is crucial. Indeed, as we shall show in the Section~\ref{estimation} [in
particular, see Eq.~(\ref{q7})], the absence of $c_{2}^{(0)}$ precludes the
possibility of identifying the DDF, given by Eq.~(\ref{ddf7}), with the one
estimated by the neoclassical theory for collisional tokamak-plasmas (see,
for example, Ref.~\cite{balescu2}). In addition, it allows describing more
complex physical scenarios such as, for example, the \textit{modified bi-Maxwelian distribution function}. Last and not least, in some
physical circumstances, the presence of $c_{2}^{(0)}$ is essential to ensure
the normalization of the DDF. Thermodynamics has been able to determine the
shape of the DDF, but it is unable to fix the seven parameters $\Theta
,\gamma ,P_{\phi 0},\lambda _{0},\Delta P_{\phi },\Delta \lambda _{0},\Delta
\lambda _{1}$. These coefficients have to be calculated in the usual way by
kinetic theory. Figs~(\ref{FDD_P_l_s})-(\ref{FDD_P_l_c}), (\ref{FDD_w_P_s})-(
\ref{FDD_w_P_c}), and (\ref{FDD_w_l_s})-(\ref{FDD_w_l_c}) illustrate three
plots of Eq.~(\ref{ddf7}) (estimated for unit values of the Jacobian and the
normalization coefficient) corresponding to the values $w=E\Theta $, $\gamma
=1+E$ (with $E$ denoting the Euler number), $P_{\phi }=P_{\phi 0}$ and $
\lambda =\lambda _{0}$, respectively. 
\begin{figure*}[tbp]
\hfill 
\begin{minipage}[t]{.50\textwidth}
    \begin{center}  
\hspace{-1.2cm}
\resizebox{1\textwidth}{!}{%
\includegraphics{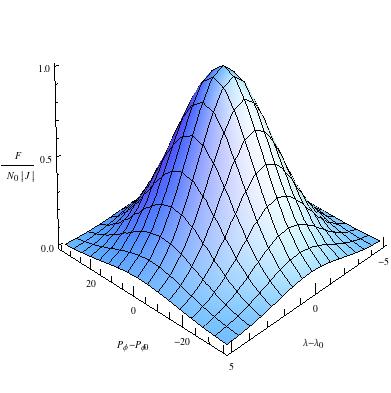}
}
\caption{ \label{FDD_P_l_s} Distribution function, Eq.~(\ref{ddf7}), (per unit values of the Jacobian and the normalization factor) computed at  $\gamma=1+E$ ($E$ indicates the Euler number), $w=E\Theta$, $\Delta P_\phi=22.360$, $\Delta\lambda_0=50.00$ and $\Delta\lambda_1=30.2031$.}
\end{center}
  \end{minipage}
\hfill 
\begin{minipage}[t]{.35\textwidth}
    \begin{center}
\hspace{-0.7cm}
\resizebox{1\textwidth}{!}{%
\includegraphics{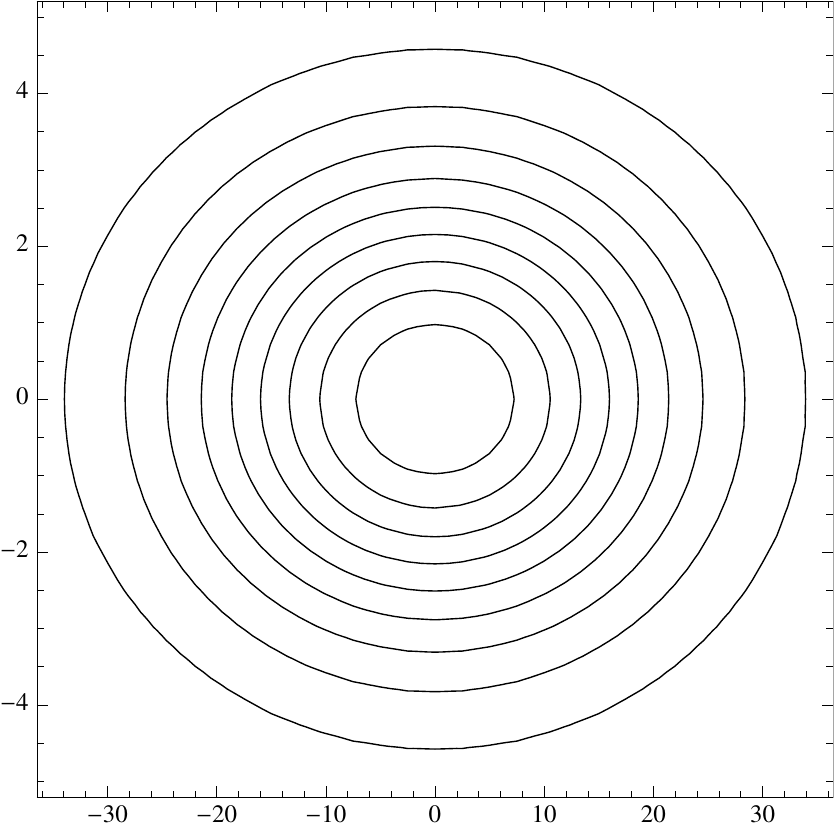}
}
\caption{Contour plot of Fig.~\ref{FDD_P_l_s}.}
\label{FDD_P_l_c}
\end{center}
  \end{minipage}
\hfill
\end{figure*}
\begin{figure*}[tbp]
\hfill 
\begin{minipage}[t]{.50\textwidth}
    \begin{center}  
\hspace{-1.2cm}
\resizebox{1\textwidth}{!}{%
\includegraphics{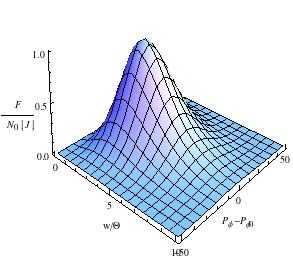}
}
\caption{ \label{FDD_w_P_s} Distribution function, Eq.~(\ref{ddf7}), (per unit values of the Jacobian and the normalization factor) computed at $\gamma=1+E$, $\Delta P_\phi=22.360$ and $\lambda=\lambda_0$.}
\end{center}
  \end{minipage}
\hfill 
\begin{minipage}[t]{.35\textwidth}
    \begin{center}
\hspace{-0.7cm}
\resizebox{1\textwidth}{!}{%
\includegraphics{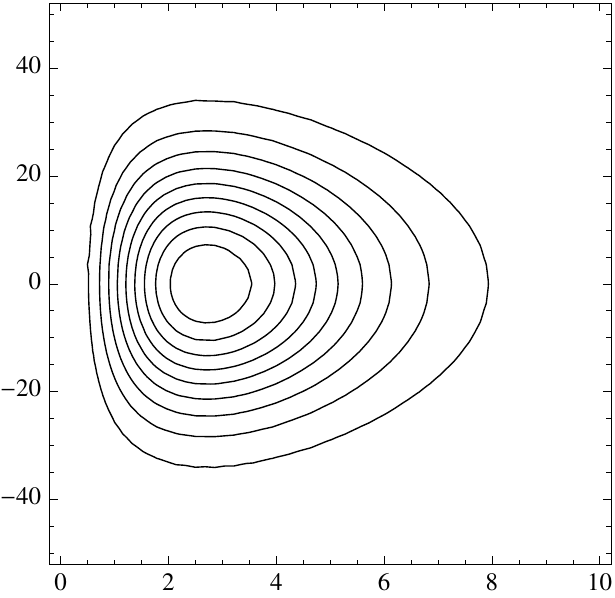}
}
\caption{Contour plot of Fig.~\ref{FDD_w_P_s}.}
\label{FDD_w_P_c}
\end{center}
  \end{minipage}
\hfill
\end{figure*}
\begin{figure*}[tbp]
\hfill 
\begin{minipage}[t]{.50\textwidth}
    \begin{center}  
\hspace{-1.2cm}
\resizebox{1\textwidth}{!}{%
\includegraphics{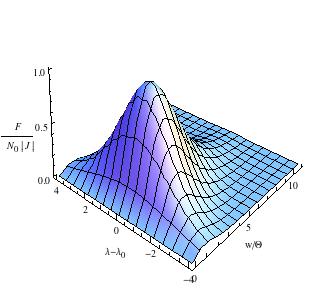}
}
\caption{ \label{FDD_w_l_s} Distribution function, Eq.~(\ref{ddf7}), (with the Jacobian and the normalization factor set to $1$) computed at $\gamma=1+E$, $P_\phi=P_{\phi0}$, $\Delta\lambda_0=11.111$ and $\Delta\lambda_1=50.00$.}
\end{center}
  \end{minipage}
\hfill 
\begin{minipage}[t]{.35\textwidth}
    \begin{center}
\hspace{-0.7cm}
\resizebox{1\textwidth}{!}{%
\includegraphics{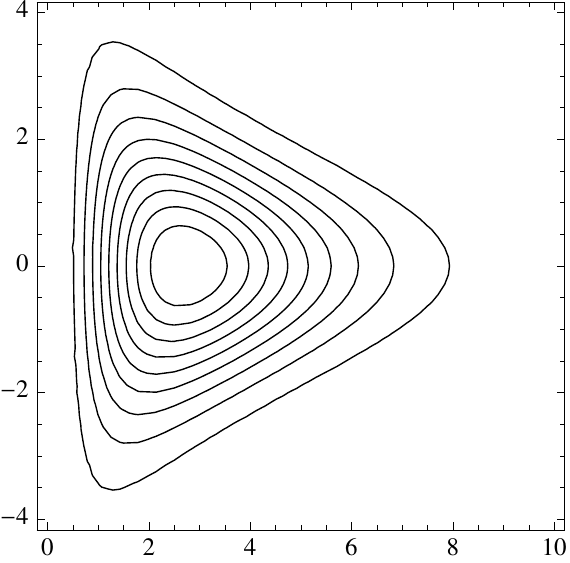}
}
\caption{Contour plot of Fig.~\ref{FDD_w_l_s}.}
\label{FDD_w_l_c}
\end{center}
  \end{minipage}
\hfill
\end{figure*}
\vskip0.2truecm
\section{Determination of the Parameters}\label{parameter} 
\vskip 0.2truecm 
\noindent This section is devoted to the
determination of the value of the parameters appearing in Eq.~(\ref{ddf8})
for ${\mathcal{F}}^R$. As an example of application, we analyze
tokamak-plasmas in collisional transport regimes. To accomplish this task we
should
\begin{itemize}
\item {Adopt a model for tokamak-plasmas};
\item {Model the source terms.}
\end{itemize}
\noindent Calculations should be performed by a kinetic approach. Representation (\ref{ddf7}) should be constructed in such a way
that the particle density $n_{\alpha }$, the average velocity $\mathbf{u}_{\alpha }$ and the temperature $T_{\alpha }$, of each species $\alpha $ (with $\alpha =e$ for
electrons and $\alpha =i$ for ions), entering in the definition of the reference state do coincide with the exact values of the density, velocity and temperature of each species. This implies (see, for example, \cite{balescu2})
\begin{equation}\label{p1a}
n_{\alpha }(\mathbf{x})=\int_{\mathcal{V}}d\mathbf{v}{\mathcal{F}}^{\alpha
R}(\mathbf{v},\mathbf{x})  
\end{equation}
\begin{equation}\label{p1b}
n_{\alpha }(\mathbf{x})\mathbf{u}^{\alpha }(\mathbf{x})=\int_{\mathcal{V}}d
\mathbf{v}\ \mathbf{v}{\mathcal{F}}^{\alpha R}(\mathbf{v},\mathbf{x}) 
\end{equation}
\noindent and
\begin{equation} \label{p1c}
n_{\alpha }(\mathbf{x})T_{\alpha }(\mathbf{x})=\frac{1}{2}m_{\alpha }\int_{
\mathcal{V}}d\mathbf{v}\mid \mathbf{v}-\mathbf{u}_{\alpha }\mid ^{2}{
\mathcal{F}}^{\alpha R}(\mathbf{v},\mathbf{x}) 
\end{equation}
\noindent where ${\mathcal{V}}$ is the velocity-volume in the phase-space and $m_{\alpha }$ is the mass particle of specie $\alpha $ [with $\alpha$=(e,i)].
To these equations we should add the entropy balance equation. The total entropy, $s$, is
defined as \cite{balescu1} 
\begin{equation}
n_{\alpha }(\mathbf{x})s_{\alpha }(\mathbf{x})=-\int_{\mathcal{V}}d\mathbf{v}
{\mathcal{F}}^{R}(\mathbf{v},\mathbf{x})\ln \Bigl(\frac{h^{3}}{em_{\alpha
}^{3}}{\mathcal{F}}^{R}(\mathbf{v})\Bigr)=-\int_{\mathcal{V}}d\mathbf{v}{
\mathcal{F}}^{R}(\mathbf{v},\mathbf{x})\ln {\mathcal{F}}^{R}(\mathbf{v})
\label{p4}
\end{equation}
\noindent Here $h$ is the Planck constant, $e$ is the (positive) magnitude
of the electric charge and $m_{\alpha }$ the mass of the species $\alpha $.
This equation is, however, inconvenient for setting the values of the
parameters, because it involves the total entropy density which, at this
stage, is not a known quantity. It is more useful to study separately the
two entropy contributions, $d_{E}S^{\alpha }$ and $d_{I}S^{\alpha }$, of the
total entropy $S^{\alpha }$: 
\begin{equation}
dS^{\alpha }=d_{E}S^{\alpha }+d_{I}S^{\alpha }\qquad \mathrm{with}\quad
S^{\alpha }=\int_{\Omega }d\mathbf{x}\ n_{\alpha }(\mathbf{x})s_{\alpha }(%
\mathbf{x})  \label{p5}
\end{equation}%
\noindent with $\Omega$ denoting the plasma-volume. $d_{E}S^{\alpha }$ represents the amounts of entropy crossing the
boundaries. The total entropy flux of species $\alpha $ is given by two
contributions: the \textit{convective entropy flux}, $n_{\alpha }s_{\alpha }%
\mathbf{u}^{\alpha }$ and the \textit{conductive entropy flux}, $\mathbf{J}%
_{S}^{\alpha }$. On the contrary, $d_{I}S^{\alpha }$ represents the source
entropy, due to internal processes, which remains within the system. In
terms of the conductive entropy flux, $\mathbf{J}_{S}^{\alpha }$, and the
entropy source strength, $\sigma ^{\alpha }$, Eq.~(\ref{p5}) takes the form 
\begin{equation}
\frac{dS^{\alpha }}{dt}=-\int_{\Sigma }d\mathbf{A}\cdot \lbrack n_{\alpha
}s_{\alpha }\mathbf{u}^{\alpha }+\mathbf{J}_{S}^{\alpha }(\mathbf{x}
)]+\int_{\Omega }d\mathbf{x}\ \sigma ^{\alpha }(\mathbf{x})  \label{p6}
\end{equation}
\noindent where $\Sigma $ denotes the boundary of the plasma-volume. According to the kinetic
theory we have \cite{degroot} 
\begin{align}
\mathbf{J}_{S}^{\alpha }(\mathbf{x})& =-\int_{\mathcal{V}}d\mathbf{v}\ [
\mathbf{v}-\mathbf{u}^{\alpha }(\mathbf{x})]{\mathcal{F}}^{\alpha R}(\mathbf{
v},\mathbf{x})\ln \Bigl(\frac{h^{3}}{em_{\alpha }^{3}}{\mathcal{F}}^{\alpha
R}(\mathbf{v},\mathbf{x})\Bigr)  \notag  \label{p7} \\
& =-\int_{\mathcal{V}}d\mathbf{v}\ [\mathbf{v}-\mathbf{u}^{\alpha }(\mathbf{x
})]{\mathcal{F}}^{\alpha R}(\mathbf{v},\mathbf{x})\ln {\mathcal{F}}^{\alpha
R}(\mathbf{v},\mathbf{x})
\end{align}
\noindent and 
\begin{equation} \label{p8}
\sigma ^{\alpha }=\frac{n_{\alpha }}{\tau _{\alpha }}\Delta _{I}S^{\alpha
}=-\sum_{\beta =e,i}\int_{\mathcal{V}}d\mathbf{v}\ [\ln {\mathcal{F}}
^{\alpha R}(\mathbf{v},\mathbf{x})]{\mathcal{K}}^{\alpha \beta } 
\end{equation}
\noindent with ${\mathcal{K}}^{\alpha \beta }$ denoting the collisional
operator of species $\alpha $ due to $\beta$, and $\tau_\alpha$ the collision time of species $\alpha$ \cite{balescu2}. We consider three kind of external energetic sources: the total Ohmic heat supplied from outside, the energy supplied by Neutral Beam Injection (NBI), and the energy of the minority population heated by Ion Cyclotron Resonant Heating (ICRH). Here, we shall not deal with burning
fusion plasmas and the loss due to the Bremsstrahlung effect is neglected. In this case, from Eq.~(\ref{p7}), we obtain 
\begin{equation} \label{p9}
-\int_{\mathcal{V}}d\mathbf{v}\ [\mathbf{v}-\mathbf{u}^{\alpha }(\mathbf{x})]
{\mathcal{F}}^{\alpha R}(\mathbf{v},\mathbf{x})\ln {\mathcal{F}}^{\alpha R}(
\mathbf{v},\mathbf{x})=\frac{1}{T_{\alpha }}(\mathbf{J}_{{\mathcal{E}}_{L}}-
\mathbf{J}_{{\mathcal{E}}_{Oh.}}-\mathbf{J}_{{\mathcal{E}}_{NBI}}-\mathbf{J}
_{{\mathcal{E}}_{ICRH}}) 
\end{equation}
\noindent Here $\mathbf{J}_{{\mathcal{E}}_{L}}$ indicates the energy loss
flux, and $\mathbf{J}_{{\mathcal{E}}_{Oh.}}$, $\mathbf{J}_{{\mathcal{E}}_{NBI}}$
and $\mathbf{J}_{{\mathcal{E}}_{ICRH}}$ denote the Ohmic energy flux, the
NBI energy flux and the ICRH energy flux, respectively. 

\noindent Eqs~(\ref{p8}) and (\ref{p9}),
together with the definitions given by Eqs~(\ref{p1a})-(\ref{p1c}),
allow determining the seven parameters $\Theta ,\gamma ,P_{\phi 0},\lambda
_{0},\Delta P_{\phi },\Delta \lambda _{0},\Delta \lambda _{1}$. More
particularly, the values of these parameters may be fixed by adopting the
following strategy. The coefficients $\gamma $, $\Delta P_{\phi },\Delta
\lambda _{0}$ and $\Delta \lambda _{1}$ are provided by Eq.~(\ref{p8}),
whereas Eq.~(\ref{p9}) sets up the calculations for the determination of
parameter $\Theta $. As we shall see in Section \ref{estimation}, $P_{\phi 0}$ and $\lambda _{0}$ are determined by
taking the limit of $P_{\phi}$ and $\lambda$ for $v_\parallel\rightarrow 0$ [see Eqs~(\ref{q1})]. The plasma temperature profile can be obtained by
Eq.~(\ref{p9}). This solution should be in agreement with the definition
given by Eq.~(\ref{p1c}). Generally, the latter calculation is very complex,
but it can be strongly simplified by adopting the following procedure. In a
first phase, the plasma is heated by the externally supplied power. In our simplified case, the applied sources are
the sum of the Ohmic, NBI and ICHR sources. As previously mentioned, calculations can be
performed only after having modeled the external energy sources and
estimated the entropy source strength $\sigma ^{\alpha }$ by modeling the
magnetically confined plasmas. Below, we provide some examples of modeling. 
\vskip0.2cm
 \noindent \textbf{The Ohmic Heating Model}
  \vskip0.2cm The
expression of the Ohmic heating density can be found in many reference
books. For the magnetic configuration Eq.~(\ref{p10a}), we have (see, for
example, Refs~\cite{balescu2}) 
\begin{align}
& {\mathcal{P}}_{Oh.}=\eta j^{2}=2\eta \Bigl[\frac{c^{2}B_{0}}{4\pi
R_{0}(1+r/R_{0}\cos \theta )}\Bigr]^{2}\frac{1}{q(0)[q(a)-1/2q(0)]}
\label{p18} \\
& \eta =\frac{\eta _{S}}{[(1-r/R_{0})^{1/2}]^{2}}
\end{align}%
\noindent with $\eta $ and $\eta _{S}$ denoting the resistivity of the plasma and the Spitzer resistivity respectively, and $\mathbf{j}$ is
the current density. $R_0$ and $a$ and the major and the minor radii of the tokamak, respectively.
\vskip0.2cm \noindent 
\textbf{The NBI Source Model} 
\vskip0.2cm Once the neutral beam enters in
the plasma, the neutral particles will be ionized. Their energy passes to
the particles of the plasmas causing heating of both electrons and ions. The
power supplied by neutral beam injection, ${\mathcal P}_{{\mathcal{E}}_{NBI}}$, may be modeled thinking in terms
of a pencil beam source, $J_{NBI}$, (see, for example, \cite{sonninocpp})
\begin{equation}
{\mathcal P}_{{\mathcal{E}}_{NBI}}={\dot{n}}_{b}\delta (\mathbf{v}-\mathbf{v}_{b})  \label{p19}
\end{equation}%
\noindent where ${\dot{n}}_{NBI}$ is the birth rate per unit volume and $%
\mathbf{v}_{b}$ is the beam ion velocity. Hence, the $NBI$ power is 
\begin{equation}
{\mathcal P}_{{\mathcal{E}}_{NBI}}\simeq \frac{1}{2}m_{b}{\dot{n}}_{b}\mid 
\mathbf{v}_{b}\mid ^{2}\Omega  \label{p20}
\end{equation}%
\noindent with $m_{b}$ denoting the mass of the beam ion. 
\vskip0.2cm 
\noindent \textbf{The ICRH Source Model} 
\vskip0.2cm 
Let us consider the case of a low concentration of ions ${\ \!{}^3\!He}$ colliding with a thermal background plasma, composed by Deuterium and electrons. The ${\ \!{}^3\!He}$ minority is about $2\%-3\%$ of the density of the background plasma and it is heated by ion cyclotron resonant heating (ICRH). In the velocity space, the evolution of the distribution function is given by the following quasi-linear balance equation \cite{brambilla} - \cite{cardinali} [see also Appendix (\ref{icrh})]
\begin{equation}\label{cm1}
\frac{\partial{\mathcal F}^m({\bf y},t)}{\partial t}=-\nabla\cdot {\bf S}^{m}({\mathcal F}^m)+P({\mathcal F}^m)
\end{equation}
\noindent where
\begin{equation}\label{cm2}
{\bf S}^{m}({\mathcal F}^m)={\bar{\bf S}}_W^{m}({\mathcal F}^m)+\sum_{\alpha=e,i}{\bar{\bf S}}_c^{m\alpha}({\mathcal F}^m)
\end{equation}
\noindent Suffices $m$ and $\alpha$ [with $m={\ \!{}^3\!He}$ and $\alpha=(e,i)$] distinguish the minority population and the species of the background plasma, respectively. The first term in Eq.~(\ref{cm2}) describes the quasi-linear diffusion due to the resonant wave particle interactions and the second term is due to the collisional operator. $P({\mathcal F}^m)$ takes into account other auxiliary sources; in our analysis, we shall put $P({\mathcal F}^m)=0$. The gradient operator, $\nabla$, is defined as the row vector $\nabla\equiv(\partial_{v_x}, \partial_{v_y}, \partial_{v_z})$ in the velocity space, whilst $\nabla\cdot{\bar{\bar{\bf A}}}$ is the matrix multiplication between the gradient vector and the matrix ${\bar{\bar{\bf A}}}$. In this approach, the ICRH is modeled by the divergence of the quasi-linear flux ${\bar{\bf S}}_W^{m}({\mathcal F}^m)$, which is proportional to the power density of the propagating wave
\begin{equation}\label{cm3}
{\mathcal P}_{ICRH}\simeq-\frac{1}{2}m_m \int_{\{\Omega{\mathcal V}\}} d\Gamma \mid{\bf v}-{\bf u}_m\mid^2\nabla\cdot{\bar{\bf S}}_W^{m}({\mathcal F}^m)
\end{equation}
\noindent where $m_m$ and ${\bf u}_m$ are the mass and the average velocity of the minority, respectively. The expression of ${\bar{\bf S}}_W^{m}({\mathcal F}^m)$, in terms of variables $P_\phi$ and $\lambda$, can be found Appendix (\ref{icrh}), Eqs~(\ref{new1}) and (\ref{new6}).
\vskip 0.2truecm 
\noindent \textbf{The Power Loss} 
\vskip0.2cm Neglecting the Bremsstrahlung loss, the rate of
energy loss is mainly due to the thermal conduction losses. Its expression
can be found in many reference books [see, for example, Refs \cite{balescu2}]). We have 
\begin{equation}
{\mathcal{P}}_{L}=3\Omega\frac{{\bar{n}}{\bar{T}}}{\tau _{E}}
\label{p21}
\end{equation}
\noindent where ${\bar{n}}$ and ${\bar{T}}$ are the average density and the
average temperature, respectively. $\tau _{E}$ is the \textit{energy
confinement time}. 
\vskip0.2cm
 \noindent \textbf{Modeling Tokamak-Plasmas} 
\vskip0.2cm The entropy source strength $\sigma ^{\alpha }$ can be estimated
by modeling the tokamak-plasma. For instance, let us consider fully ionized
tokamak-plasmas, defined as a collection of magnetically confined electrons
and positively charged ions. In the \textit{local triad} ($\mathbf{e}_{r},
\mathbf{e}_{\theta },\mathbf{e}_{\phi }$), the magnetic field, in the
standard high aspect ratio, low $beta$ (the plasma pressure normalized to
the magnetic field strength), circular tokamak equilibrium model, reads
(see, for example, Ref.~\cite{balescu2}) (see the footnote \footnote{
Note that in the limit ${\hat{\eta}}\equiv r/R_{0}\ll 1$, the ratio between
the poloidal component and the toroidal component of the magnetic field
strength is independent of the poloidal angle $\theta $ (at the first
approximation in $\hat{\eta}$).\label{mfield}}) 
\begin{equation}
\mathbf{B}=\frac{B_{0}}{q(r)}\frac{r}{R_{0}}\mathbf{e}_{\theta }+\frac{B_{0}
}{1+(r/R_{0})cos\theta }\mathbf{e}_{\phi }  \label{p10a}
\end{equation}%
\noindent Here $B_{0}$ is a constant having the dimension of a magnetic
field intensity, and $q(r)$ is the safety factor (in Ref.~\cite%
{balescu2} the reader can find an exact definition of this tokamak
parameter), respectively. In the magnetic configuration, given by Eq.~(\ref{p10a}), we have 
\begin{equation}
F=B_{0}R_{0}\qquad ;\qquad \psi (r)=2\pi B_{0}\int_{0}^{r}\frac{r}{q(r)}\ dr
\label{pq1}
\end{equation}%
\noindent The dimensionless entropy production of species $\alpha $, $\Delta
_{I}S^{\alpha }$, is derived under the sole assumption that the state of the
quiescent plasma is not too far from the reference local Maxwellian. In the 
\textit{local dynamical triad}, $\Delta _{I}S^{\alpha }$ can
be brought into the form (see Refs~\cite{balescu2} and \cite{sonnino}) 
\begin{align}\label{p11}
\Delta _{I}S^{e}=& \ q_{\parallel ps}^{(1)}(g_{\parallel }^{(1)}-{\bar{g}}%
_{\parallel }^{e(1)})+q_{\parallel ps}^{e(3)}(g_{\parallel }^{e(3)}+{\bar{g}}%
_{\parallel }^{e(3)})+q_{\parallel b}^{(1)}(g_{\parallel }^{(1)}-{\bar{g}}%
_{\parallel }^{e(1)})+q_{\parallel b}^{e(3)}(g_{\parallel }^{e(3)}+{\bar{g}}%
_{\parallel }^{e(3)})  \\
& +q_{\parallel b}^{e(5)}{\bar{g}}_{\parallel }^{e(5)}+{\hat{q}}_{\rho
cl}^{e(1)}g_{\rho }^{(1)P}+{\hat{q}}_{\rho cl}^{e(3)}g_{\rho }^{e(3)} \\
\Delta _{I}S^{i}=& \ q_{\parallel ps}^{i(3)}(g_{\parallel }^{i(3)}+{\bar{g}}%
_{\parallel }^{i(3)})\ +q_{\parallel b}^{i(3)}(g_{\parallel }^{i(3)}+{\bar{g}%
}_{\parallel }^{i(3)})+q_{\parallel b}^{i(5)}{\bar{g}}_{\parallel }^{i(5)}+{%
\hat{q}}_{\rho cl}^{i(3)}g_{\rho }^{i(3)}  \notag
\end{align}%
\noindent Here $q_{r}^{\alpha (n)}$ [with $r=$($\rho ,\parallel ,\wedge $)]
denote the \textit{Hermitian moments of the distribution functions} and $%
g_{r}^{\alpha (n)},\ {\bar{g}}_{r}^{\alpha (n)}$ are the \textit{%
dimensionless source terms}. Index $n$ takes the values $n=(1,3,5)$. In the
linear Onsager region, and up to the second order of the drift parameter $%
\epsilon $, it can be shown that Eqs~(\ref{p11}) simplify to \cite{balescu2} 
\begin{align}\label{p12}
\Delta _{I}S^{e}=& \ {\tilde{\sigma}}_{\parallel }(g_{\parallel }^{(1)}-{%
\bar{g}}_{\parallel }^{e(1)})^{2}+{\tilde{\kappa}}_{\parallel
}^{e}(g_{\parallel }^{e(3)}+{\bar{g}}_{\parallel }^{e(3)})^{2}+{\tilde{%
\epsilon}}_{\parallel }^{e}({\bar{g}}_{\parallel }^{e(5)})^{2}+2{\tilde{%
\alpha}}_{\parallel }(g_{\parallel }^{(1)}-{\bar{g}}_{\parallel
}^{e(1)})(g_{\parallel }^{e(3)}+{\bar{g}}_{\parallel }^{e(3)}) \nonumber\\
& +2{\tilde{\gamma}}_{\parallel }(g_{\parallel }^{(1)}-{\bar{g}}_{\parallel
}^{e(1)}){\bar{g}}_{\parallel }^{e(5)}+2{\tilde{\delta}}_{\parallel
}^{e}(g_{\parallel }^{e(3)}+{\bar{g}}_{\parallel }^{e(3)}){\bar{g}}%
_{\parallel }^{e(5)} \nonumber\\
& +{\tilde{\sigma}}_{\perp }(g_{\rho }^{(1)P})^{2}+{\tilde{\kappa}}_{\perp
}^{e}(g_{\rho }^{e(3)})^{2}-2{\tilde{\alpha}}_{\perp }g_{\rho
}^{(1)P}g_{\rho }^{e(3)}\nonumber\\
\Delta _{I}S^{i}=& \ {\tilde{\kappa}}_{\parallel }^{i}(g_{\parallel }^{i(3)}+%
{\bar{g}}_{\parallel }^{i(3)})^{2}+{\tilde{\epsilon}}_{\parallel }^{i}({\bar{%
g}}_{\parallel }^{i(5)})^{2}+2{\tilde{\delta}}_{\parallel }^{i}(g_{\parallel
}^{i(3)}+{\bar{g}}_{\parallel }^{i(3)}){\bar{g}}_{\parallel }^{i(5)}+{\tilde{%
\kappa}}_{\perp }^{i}(g_{\rho }^{i(3)})^{2} 
\end{align}%
\noindent where coefficients ${\tilde{\sigma}}_{r}$, ${\tilde{\alpha}}_{r}$, 
${\tilde{\kappa}}_{r}^{\alpha }$ indicate the dimensionless component of the 
\textit{electronic conductivity}, the \textit{thermoelectric coefficient}
and the \textit{electric} ($\alpha =e$) or \textit{ion} ($\alpha =i$) 
\textit{thermal conductivity}, respectively. Moreover, ${\tilde{\gamma}}%
_{\parallel }$, ${\tilde{\delta}}_{\parallel }^{\alpha }$ and ${\tilde{%
\epsilon}}_{\parallel }^{\alpha }$ are the \textit{parallel transport
coefficients} in 21M approximation. As shown in Eq.~(\ref{p8}), the entropy
production is closely associated with the collision term. Notice that Eqs.~(%
\ref{p12}) have been derived by using the following \textit{linearized
Landau collisional term} 
\begin{align}
{\mathcal{K}}_{L}^{\alpha \beta }=& \frac{2\pi e_{\alpha }^{2}e_{\beta
}^{2}\ln \Lambda }{m_{\alpha }}\int d\mathbf{v}_{2}\frac{\partial }{\partial
v_{1r}}G_{rs}(\mathbf{v}_{1}-\mathbf{v}_{2})\Bigl(f_{0}^{\beta }(\mathbf{v}%
_{2})\frac{1}{m_{\alpha }}\frac{\partial }{\partial v_{1s}}f_{1}^{\alpha }(%
\mathbf{v}_{1})  \notag  \label{p13} \\
& +f_{1}^{\beta }(\mathbf{v}_{2})\frac{1}{m_{\alpha }}\frac{\partial }{%
\partial v_{1s}}f_{0}^{\alpha }(\mathbf{v}_{1})-f_{0}^{\alpha }(\mathbf{v}%
_{1})\frac{1}{m_{\beta }}\frac{\partial }{\partial v_{2s}}f_{0}^{\beta }(%
\mathbf{v}_{2})-f_{1}^{\alpha }(\mathbf{v}_{1})\frac{1}{m_{\beta }}\frac{
\partial }{\partial v_{2s}}f_{0}^{\beta }(\mathbf{v}_{2})\Bigr)
\end{align}
\noindent Moreover, $G_{rs}(\mathbf{g})$ and $\ln \Lambda $ denote the 
\textit{Landau tensor} and the \textit{Coulomb logarithm}, respectively 
\begin{equation}
G_{rs}(\mathbf{g})=\frac{g^{2}\delta _{rs}-g_{r}g_{s}}{g^{3}}\quad ;\quad
\ln \Lambda =\frac{3/2(T_{e}+T_{i})\lambda _{D}}{Ze^{2}}\quad ;\quad \lambda
_{D}=\Bigl(\frac{4\pi Ze^{2}(n_{e}T_{e}+n_{i}T_{i})}{T_{e}T_{i}(1+Z)}\Bigr)
^{-1/2}  \label{p14}
\end{equation}
\noindent Here $Z$ is the \textit{charge number of ions}. In addition, in
Eq.~(\ref{p13}), the distribution function is expanded in powers of the
drift parameter $\epsilon $ 
\begin{align}
& f^{\alpha }(\mathbf{v})=f_{0}^{\alpha }(\mathbf{v})+\epsilon f_{1}^{\alpha
}(\mathbf{v})+\cdots \qquad \qquad \qquad \mathrm{with}  \notag  \label{p15}
\\
& f_{0}^{\alpha }(\mathbf{v})=n_{\alpha }\Bigl(\frac{m_{\alpha }}{2\pi
T_{\alpha }}\Bigr)^{3/2}\exp \Bigl(-\frac{m_{\alpha }}{2T_{\alpha }}v^{2}
\Bigr)
\end{align}
\noindent By summarizing, the expression of the entropy source strength,
Eq.~(\ref{p12}), may be put on the left-hand-side of Eq.~(\ref{p8}) provide
that the integral appearing on the right-hand-side of this equation is
evaluated by means of the linearized Landau operator Eq.~(\ref{p13}).
\vskip 0.2truecm 
\section{Estimation of the Seven Parameters Appearing in the Expression of ${\mathcal F}^R$ for the Simple Model of Fully Ionized Tokamak-Plasmas}\label{estimation}
 \vskip 0.2truecm 
\noindent In this section we show that the reference DDF, ${\mathcal{F}}^R$, coincides exactly with the one predicted by the neoclassical theory for (non turbulent) collisional tokamak-plasmas in Onsager's region. According to our formalism, from Eq.~(\ref{i1}) we see that two density distribution functions coincide if, and only if, the entropy productions  are identical {\it for all values taken by the variables}. In our case, we should check that the entropy production in Eq.~(\ref{ddf7}) can be identified with the one given by Eq.~(\ref{p12})  {\it for all values taken by the thermodynamic forces}. We shall see that this is possible and, at the same time, such an identification will allow determining the free parameters appearing in ${\mathcal{F}}^R$.
\noindent From Eqs~(\ref{i4}) and (\ref{ddf3}), we have 
\begin{equation}  \label{ex1}
\Delta_IS=-(\gamma-1)\ln\Bigl(\frac{w}{\Theta}\Bigr)+\frac{w}{\Theta}+\frac{1
}{2}g_{11}\alpha_1^2 +\frac{1}{2}g_{22}\alpha_2^2+g_{12}\alpha_1\alpha_2
\end{equation}
\noindent By expanding the previous expression around the reference value $
w=w_0$ we obtain, up to the second order 
\begin{align}  \label{ex2}
\Delta_IS=&\ \Bigl[-(\gamma-1)\ln\Bigl(\frac{w_0}{\Theta}\Bigr)+\frac{w_0}{
\Theta}\Bigr]+\Bigl[-(\gamma-1)\frac{1}{w_0}+\frac{1}{\Theta}\Bigr]
(w-w_0)+(\gamma-1)\frac{1}{2w_0^2}(w-w_0)^2  \notag \\
&+\frac{1}{2}g_{11}\alpha_1^2+\frac{1}{2}g_{22}\alpha_2^2+g_{12}\alpha_1
\alpha_2+h.o.t.
\end{align}
\noindent Up to a normalization constant, we have that the distribution
function Eq.~(\ref{ddf7}) is approximated by a Gaussian density distribution
function (in the variable $w$) by setting to zero the coefficient of the
linear term [i.e., $-(\gamma -1)/w_{0}+1/\Theta =0$]. The \textit{global
optimality conditions} is obtained by imposing that also the sum of the
constant terms vanishes [i.e., $-(\gamma -1)\ln(w_{0}/\Theta)+w_0/\Theta =0$
]. These two requirements are simultaneously satisfied only if 
\begin{align}  \label{ex3}
&\gamma=1+E \\
&w_0=(\gamma-1)\Theta=E\Theta  \notag
\end{align}
\noindent Hence, solutions~(\ref{ex3}) ensure not only a local approximation, valid up to the
second order, but also a good global approximation. In this sense the values
of $w_{0}$ and $\gamma $, provided by Eq.~(\ref{ex3}), are \textit{optimal}.
From Eq.~(\ref{i2}) and Eq.~(\ref{ex2}) we obtain the expressions of the
thermodynamic forces 
\begin{align}  \label{ex4}
&X^1\equiv\frac{\partial \Delta_IS}{\partial\alpha_1}=g_{11}\alpha_1+g_{12}%
\alpha_2  \notag \\
&X^2\equiv\frac{\partial \Delta_IS}{\partial\alpha_2}=g_{12}\alpha_1+g_{22}%
\alpha_2 \\
&X^3\equiv\frac{\partial \Delta_IS}{\partial w}=\frac{1}{E\Theta^2}(w-w_0) 
\notag
\end{align}
\noindent By solving the previous system of equations with respect to $%
\alpha_1$, $\alpha_2$ and ($w-w_0$), we find 
\begin{align}  \label{ex5}
&\alpha_1={\hat g}_{22}X^1-{\hat g}_{12}X^2  \notag \\
&\alpha_2=-{\hat g}_{12}X^1+{\hat g}_{11}X^2\quad \\
&(w-w_0)=E\Theta^2 X^3\qquad\mathrm{where}  \notag \\
&{\hat g}_{j\kappa}\equiv\frac{g_{j\kappa}}{g}\quad [\mathrm{with}\
(j,\kappa)=(1,2)]\qquad ; \quad g\equiv g_{11}g_{22}-g_{12}^2=\frac{1}{{\hat
g}_{11}{\hat g}_{22}-{\hat g}_{12}^2}=\frac{1}{{\hat g}}  \notag
\end{align}
\noindent Hence, in terms of the thermodynamic forces, the electron and ion
entropy source strength read, respectively 
\begin{align}  \label{ex6}
&\Delta_IS^e=\frac{1}{2}E\Theta_e^2{X_e^{3}}^2+\frac{1}{2}{\hat g}_{22}^e{%
X_e^{1}}^2+\frac{1}{2}{\hat g}_{11}^e{X_e^{2}}^2-{\hat g}%
_{12}^eX_e^{1}X_e^{2}+h.o.t. \\
&\Delta_IS^i=\frac{1}{2}E\Theta_i^2{X_i^{2}}^2+\frac{1}{2}{\hat g}_{11}^i{%
X_i^{1}}^2+h.o.t.  \notag
\end{align}
\noindent After diagonalization, the first expression of Eq.~(\ref{ex6}) can be
brought into the form 
\begin{align}  \label{ex6a}
&\Delta_IS^e=\frac{1}{2}E\Theta_e^2{X_e^{3}}^2+\mu_1{\xi_e^1}^2+\mu_2{\xi_e^2%
}^2\qquad\quad \mathrm{with} \\
&\mu_{1,2}=\frac{1}{4}\Bigl[({\hat g}_{11}+{\hat g}_{22})\pm\sqrt{({\hat g}%
_{11}-{\hat g}_{22})^2+4{\hat g}_{12}^2}\ \Bigr]  \notag \\
&\xi_e^1=\frac{q_1}{h}{\hat g}_{12}X_e^{1}-\frac{q_1}{2h}({\hat g}_{11}-{%
\hat g}_{22}-h)X^{2}_e  \notag \\
&\xi_e^2=-\frac{q_2}{h}{\hat g}_{12}X_e^{1}+\frac{q_2}{2h}({\hat g}_{11}-{%
\hat g}_{22}+h)X^{2}_e  \notag \\
&h=\sqrt{({\hat g}_{11}-{\hat g}_{22})^2+4{\hat g}_{12}^2}  \notag \\
&q_{1,2}=\frac{\sqrt{4{\hat g}_{12}^2+({\hat g}_{11}-{\hat g}_{22})({\hat g}%
_{11}-{\hat g}_{22}\pm h)}}{\mid {\hat g}_{12}\mid\sqrt{2}}
\end{align}
\noindent In the banana regime, without considering
the classical contributions, the expressions of the electron and ion entropy
source strength take the form \cite{balescu2}
\begin{align}  \label{ex7}
{\Delta_I S^e}{{}^{}}^\prime= &\ {\tilde\epsilon}_\parallel^e{Z_e^1}^2+{%
\tilde\kappa}^e_\parallel{Z_e^2}^2+{\tilde\sigma}_\parallel{Z_e^3}^2+2{%
\tilde\delta}_\parallel^eZ_e^1Z_e^2+2{\tilde\gamma}_\parallel Z_e^1Z_e^3+ 2{%
\tilde\alpha}_\parallel Z_e^2Z_e^3  \notag \\
{\Delta_I S^i}{{}^{}}^\prime= &\ {\tilde\epsilon}^i_\parallel {Z_i^1}^2+{%
\tilde\kappa}_\parallel^i{Z_i^2}^2+2{\tilde\delta}^i_\parallel Z_i^1z_i^2
\end{align}
\noindent where 
\begin{equation}  \label{ex8}
\left\{ 
\begin{array}{ll}
Z_e^1={\bar g}_\parallel^{e(5)} & \quad\mbox{$ $} \\ 
Z_e^2=g_\parallel^{e(3)}+{\bar g}_\parallel^{e(3)} & \quad\mbox{ $$} \\ 
Z_e^3\equiv g_\parallel^{(1)}-{\bar g}_\parallel^{e(1)} & \quad\mbox{ $$}%
\end{array}
\right. \qquad ;\qquad\ \ \left\{ 
\begin{array}{ll}
Z_i^1=g_\parallel^{i(3)}+{\bar g}_\parallel^{i(3)} & \quad\mbox{$ $} \\ 
Z_i^2={\bar g}_\parallel^{i(5)} & \quad\mbox{ $$}
\end{array}
\right.
\end{equation}
\noindent As mentioned in the introduction [see Eq.~(\ref{i3})], thermodynamic systems obtained by a transformation of forces and fluxes in such a way that the entropy production remains unaltered are thermodynamically equivalent [Thermodynamic Covariance Principle (TCP)] \cite{prigogine}, \cite{sonninoPRE}. The following linear transformation of the forces 
\begin{equation}  \label{ex9}
\left\{ 
\begin{array}{ll}
X_e^1=Z_e^1={\bar g}_\parallel^{e(5)} & \quad\mbox{$ $} \\ 
X_e^2=Z_e^2=g_\parallel^{e(3)}+{\bar g}_\parallel^{e(3)} & \quad\mbox{ $$}
\\ 
X_e^3\equiv a_e Z_e^1 +b_e Z_e^2 +c_e Z_e^3 & \quad\mbox{ $$}
\end{array}
\right. \qquad ;\qquad\ \ \left\{ 
\begin{array}{ll}
X_i^1=Z_i^1=g_\parallel^{i(3)}+{\bar g}_\parallel^{i(3)} & \quad\mbox{$ $}
\\ 
X_i^2=a_iZ_i^1+b_iZ_i^2 & \quad\mbox{ $$}
\end{array}
\right.
\end{equation}
\noindent leaves unaltered the entropy production : $\Delta_I S^\alpha={\Delta_I S^\alpha}{{}^{}}^\prime$ with $\alpha=(e,i)$. Hence, the reference DDF given by Eq.~(\ref{ddf7}) coincides with the reference distribution function estimated by the neoclassical theory if, and only if, the expressions of the entropy productions given by Eqs~(\ref{ex7}) are identical to Eqs~(\ref{p12}). By imposing the validity of these identities, we get the expressions of the transport coefficients ${\hat g}^\alpha_{ij}$ and the transformation-coefficients $a_\alpha$, $b_\alpha$ [with $\alpha=(e,i)$] and $c_e$
\begin{align}  \label{ex11}
&a_e=\frac{2{\tilde\gamma}_\parallel}{\sqrt{E{\tilde\sigma}_\parallel}%
\Theta_e}\quad ;\quad b_e=\frac{2{\tilde\alpha}_\parallel }{\sqrt{E{%
\tilde\sigma}_\parallel}\Theta_e}\quad ;\quad c_e=\frac{2\sqrt{{\tilde\sigma}%
_\parallel}}{\sqrt{E}\Theta_e}  \notag \\
& {\hat g}^e_{11}=\frac{2}{{\tilde\sigma}_\parallel}({\tilde\kappa}%
^e_\parallel {\tilde\sigma}_\parallel-{\tilde\alpha}_\parallel ^2)\quad
;\quad {\hat g}^e_{22}=\frac{2}{{\tilde\sigma}_\parallel}({\tilde\epsilon}%
_\parallel^e {\tilde\sigma}_\parallel-{\tilde\gamma}_\parallel ^2)\quad
;\quad {\hat g}^e_{12}=\frac{2}{{\tilde\sigma}_\parallel}({\tilde\alpha}%
_\parallel{\tilde\gamma}_\parallel-{\tilde\delta}_\parallel^e {\tilde\sigma}%
_\parallel) \\
&a_i=\frac{2{\tilde\delta}^i_\parallel}{\sqrt{E{\tilde\kappa}_\parallel^i}%
\Theta_i}\quad;\quad b_i=\frac{2\sqrt{{\tilde\kappa}_\parallel^i}}{\sqrt{E}%
\Theta_i}\quad\ ;\quad {\hat g}^i_{11}=\frac{2}{{\tilde\kappa}_\parallel^i}({%
\tilde\epsilon}_\parallel^i {\tilde\kappa}_\parallel^i-{\tilde\delta}^i_%
\parallel{}^2)  \notag
\end{align}
\noindent To sum up, the set of Eqs~(\ref{ex11}) ensures that for collisional tokamak-plasmas in the Onsager region, the reference density distribution function given by Eq.~(\ref{ddf7}) identifies with the reference DDF estimated by the neoclassical theory.  

\noindent By inserting the values of the collision matrix elements (see, for
example, Ref.~\cite{balescu1}), we find 
\begin{align}  \label{newex}
&a_e=\frac{0.1168}{\Theta_e}\quad ;\quad b_e=-\frac{0.7599}{\Theta_e}\quad
;\quad \ \ c_e=\frac{1.6943}{\Theta_e}  \notag \\
& {\hat g}^e_{11}=2.5125\quad ;\quad {\hat g}^e_{22}=1.9090\quad\ ;\quad\ \ {%
\hat g}^e_{12}=-1.6583 \\
&a_i=\frac{0.6364}{\Theta_i}\quad\ ;\quad b_i=\frac{1.8033}{\Theta_i}\quad\
\ ;\quad\ \ {\hat g}^i_{11}=2.1994  \notag
\end{align}
\noindent The numerical values of parameters $\Theta_\alpha$ can be obtained
by means of Eqs~(\ref{p1a}), (\ref{p1b}) and (\ref{p9}) where the
distribution functions ${\mathcal{F}}^{\alpha0}$ are constructed by using
Eqs~(\ref{ex6}). From Eq.~(\ref{p1a}) we obtain an equation, depending on
variable $\Theta_\alpha$, for the particle density $n_\alpha$. 
\begin{equation}  \label{app1}
n_\alpha (\mathbf{x,\Theta_\alpha})=\int d\mathbf{v} {\mathcal{F}}^{\alpha
0}(\Theta_\alpha, \mathbf{v},\mathbf{x})\qquad ; \quad\alpha=(e,i)
\end{equation}
\noindent By injecting Eq.~(\ref{app1}) into Eq.~(\ref{p1b}), we have 
\begin{equation}  \label{app2}
\mathbf{u}_\alpha(\mathbf{x},\Theta_\alpha) =\frac{\int d\mathbf{v} \mathbf{v%
}{\mathcal{F}}^{\alpha 0}(\Theta_\alpha, \mathbf{v},\mathbf{x})}{\int d%
\mathbf{v} {\mathcal{F}}^{\alpha 0}(\Theta_\alpha, \mathbf{v},\mathbf{x})}
\end{equation}
\noindent This latter equation should then be combined with Eq.~(\ref{p9})
and, for a given value of $\mathbf{J}_{{\mathcal{E}}_{Oh.}}$, we finally derive the
equation for $\Theta_\alpha$ 
\begin{equation}\label{app3}
\int_{\Omega\mathcal{V}}d\mathbf{v}d{\bf x}\ [\mathbf{v}-\mathbf{u}
_\alpha(\mathbf{x},\Theta_\alpha)]
{\mathcal{F}}^{\alpha R}(\mathbf{v},\mathbf{x})\ln {\mathcal{F}}^{\alpha R}
(\mathbf{v},\mathbf{x},\Theta_\alpha)=\int_\Omega d{\bf x}\ \frac{\mathbf{J}_{{\mathcal{E}}_{Oh.}}}{T_{\alpha }}
\qquad ;
\quad\alpha=(e,i)
\end{equation}
\noindent with $T_\alpha$ provided by Eq.~(\ref{p1c}). Eq.~(\ref
{app3}) can be solved (numerically) with respect to variable $\Theta_\alpha$%
; we find 
\begin{equation}  \label{t2}
(\Theta_e,\Theta_i)=(3.2970\times 10^{18} cm^2sec^{-2},1.9325\times
10^{15}cm^2sec^{-2})
\end{equation}
\noindent From Eq.~(\ref{app1}), Eq.~(\ref{app2}) and Eq.~(\ref{p1c}), we
can now derive the profiles of the particle density, $n_\alpha(\mathbf{x})$,
the average velocity, $\mathbf{u}_\alpha$, and temperature, $T_\alpha$,
respectively. The knowledge of the fields $n_\alpha$, $T_\alpha$ etc. allows
determining the profiles of the thermodynamic forces $Z^j_\alpha$ [j=(1,2,3)
for electrons, and j=(1,2) for ions]. Parameters $P_{\phi 0}$ and $\lambda_0$
are determined in the following manner. We note that by substituting Eqs~(\ref{ddf1}) into Eq.~(\ref{ddf7}), the distribution function, Eq.~(\ref
{ddf7}) tends to reduce to a pure gamma-process as $P_\phi\rightarrow
\psi(r)$, $\lambda\rightarrow (2\!\mid\!\!B\!\!\mid)^{-1}$ and $
v_\parallel\rightarrow 0$ ( i.e., the pitch angle $\theta_P$ is close to $
\pi/2$). This happens when the thermodynamic forces, $X^1$ and $X^2$, tend
to vanish. Our aim is to derive the expressions of the steady-state electron
and ion distribution functions for plasmas in the Onsager region, and $
v_\parallel$, $w$, $P_\phi$ and $\lambda$ close to $0$, $E\Theta$, $\psi(r)$
and $[2\!\mid\!\! B\!\!\mid\!\!(r,\theta)\ \!]^{-1}\!\!\!\!\!\mathrm{,}\ \
\! $ respectively (see also the footnote \footnote{In our calculations, we shall assume that, other than $v_\parallel$, also $%
dv_\parallel$ is close to $0$.}). To this purpose, we shall work with the
dimensionless variables ${\hat P}_\phi$ and $\hat\lambda$, which for the
magnetic configuration given by Eq.~(\ref{p10a}), are defined as 
\begin{align}  \label{qn1}
&{\hat P}_\phi\equiv\frac{1}{B_0a^2}P_\phi=\frac{B_0R_0}{a^2\Omega_{0c}}\frac{%
v_\parallel}{\mid B\mid}+2\pi\int_0^\rho\frac{\rho^{\prime }}{q(\rho^{\prime
})}\ d\rho^{\prime } \\
&{\hat\lambda}\equiv B_0\lambda=\frac{B_0}{2\mid B\mid}\Bigl(1-\frac{%
v_\parallel^2}{w}\Bigr)  \notag \\
&\mathrm{with}\quad \rho\equiv\frac{r}{a}\qquad ;\qquad \mid B\mid\simeq B_0%
\Bigl(1-\frac{a}{R_0}\rho\cos\theta\Bigr)  \notag
\end{align}
\noindent We have 
\begin{align}  \label{q1}
&{\hat P}_{\phi 0}^\alpha={\hat P}_{\phi
}^\alpha\mid_{v_\parallel\rightarrow0}\ =2\pi\int_0^{\rho_{\alpha 0}}\frac{%
\rho^{\prime }}{q(\rho^{\prime })}\ d\rho^{\prime }={\hat \psi}(\rho_{\alpha
0})\qquad\qquad\ \! ; \quad\alpha=(e,i) \\
&{\hat\lambda}_0^\alpha={\hat\lambda}^\alpha\mid_{v_\parallel\rightarrow0}\ =%
\frac{B_0}{2\mid B\mid}\simeq \frac{1}{2}\Bigl(1+\frac{a}{R_0}\rho_{\alpha
0}\cos\theta_{\alpha 0}\Bigr)\qquad\quad\ \!\! \!; \quad \alpha=(e,i)  \notag
\end{align}
\noindent where $(\rho_{e0},\theta_{e0})$ and $(\rho_{i0},\theta_{i0})$ are
solutions of the equations (see the footnote \footnote{%
At the values $X^1_e=X^2_e=0$ and $X_i^1=0$, ${\mathcal{F}}^R_\alpha$
exactly coincides with a gamma distribution function. Since the entropy
productions, Eqs~(\ref{p12}), have been estimated by truncating the
expressions at the second order in the drift parameters $\epsilon$ \cite%
{balescu1}, for coherence, Eqs~(\ref{q2}) and (\ref{q3}) should also be
solved up to the second order in $\epsilon$. We are then imposing the
conditions for which ${\mathcal{F}}^R_\alpha$ approximates the gamma
distribution function up to $\epsilon^2$.}) 
\begin{equation}  \label{q2}
\left\{ 
\begin{array}{ll}
X^{1}_e(r,\theta)=g_\parallel^{e(3)}+{\bar g}_\parallel^{e(3)}={\mathcal{O}}%
(\epsilon) & \quad\mbox{$ $} \\ 
X^{2}_e(r,\theta)={\bar g}_\parallel^{e(5)}={\mathcal{O}}(\epsilon) & \quad%
\mbox{ $$}%
\end{array}
\right. ;\quad \left\{ 
\begin{array}{ll}
X^{1}_i(r,\theta)=g_\parallel^{i(3)}+{\bar g}_\parallel^{i(3)}={\mathcal{O}}%
(\epsilon) & \quad\mbox{$ $} \\ 
r_{i0}\in \{ r\mid x^{(1)}_i(r,\theta)={\mathcal{O}}(\epsilon)\} & \quad%
\mbox{ $$}%
\end{array}
\right.
\end{equation}
\noindent with the drift parameter $\epsilon$ of the order $\epsilon\simeq
10^{-3}$ \cite{balescu2}. As an example of calculation, we choose the
following values of the parameters: $B_0=3.45\ Tesla$, $R_0=2,96\ m$ and $%
a=1,25\ m$. In this case, the systems (\ref{q2}) admits the solutions 
\begin{equation}  \label{q3}
\left\{ 
\begin{array}{ll}
(\rho_{e0},\theta_{e0})=(0.2376, 1.0562) & \quad\mbox{$ $} \\ 
X^{1}_e=X^{2}_e=2.0367\times 10^{-4}\sim {\mathcal{O}}(\epsilon) & \quad%
\mbox{ $$}%
\end{array}
\right. ;\quad \left\{ 
\begin{array}{ll}
(\rho_{i0},\theta_{i0})=(0.5079, 2.8137) & \quad\mbox{ $$} \\ 
X_i^{1}=0 & \quad\mbox{ $$}%
\end{array}
\right.
\end{equation}
\noindent Fig.~(\ref{FDDe}) illustrates the surfaces of the electron forces $%
X^{1}_e$ and $X^{2}_e$. The intersection line corresponds to the values of $%
(\rho,\theta)$ such that $X^{1}_e=X^{2}_e$. Fig.~(\ref{FDDi}) shows the
curve $\rho=\rho(\theta)$ where the ion thermodynamic force $X_i^{1}$
vanishes. The values of parameters $P^\alpha_{\phi0}$ and $%
\lambda^\alpha_{0} $ are easily estimated from Eqs~(\ref{q1}) 
\begin{figure*}[tbp]
\hfill 
\begin{minipage}[t]{.50\textwidth}
    \begin{center}  
\hspace{-1.2cm}
\resizebox{1\textwidth}{!}{%
\includegraphics{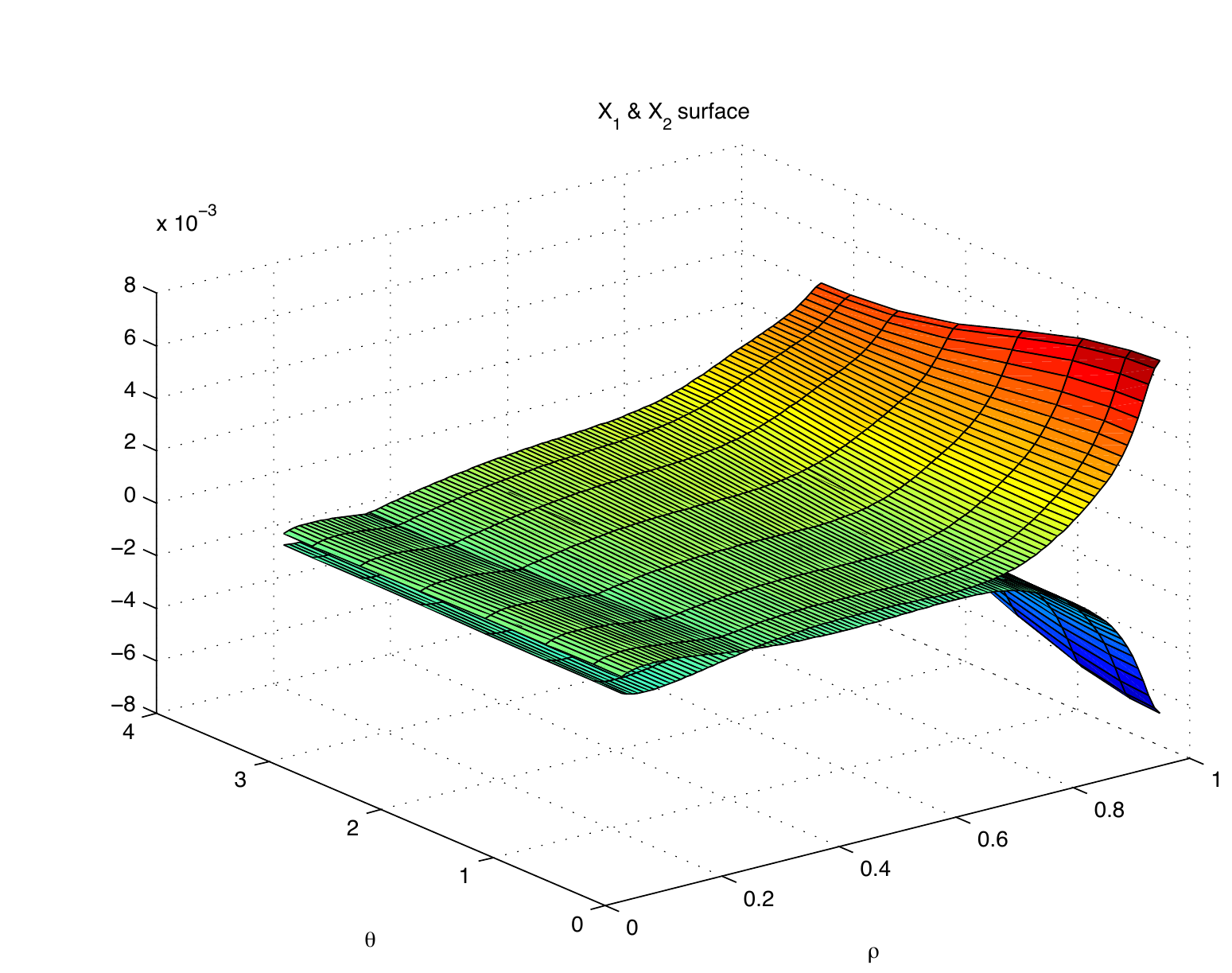}
}
\caption{ \label{FDDe} The electron thermodynamic forces $X^{1}_e$ and $X^{2}_e$. The intersection line corresponds to the values $(\rho,\theta)$ for which $X^{1}_e=X^{2}_e$. The first system of equations in Eqs~(\ref{q2}), is satisfied in the narrow region around to the values $(\rho_{e0},\theta_{e0})=(0.2376, 1.0562)$.}
\end{center}
  \end{minipage}
\hfill 
\begin{minipage}[t]{0.40\textwidth}
    \begin{center}
\hspace{-0.95cm}
\resizebox{1\textwidth}{!}{%
\includegraphics{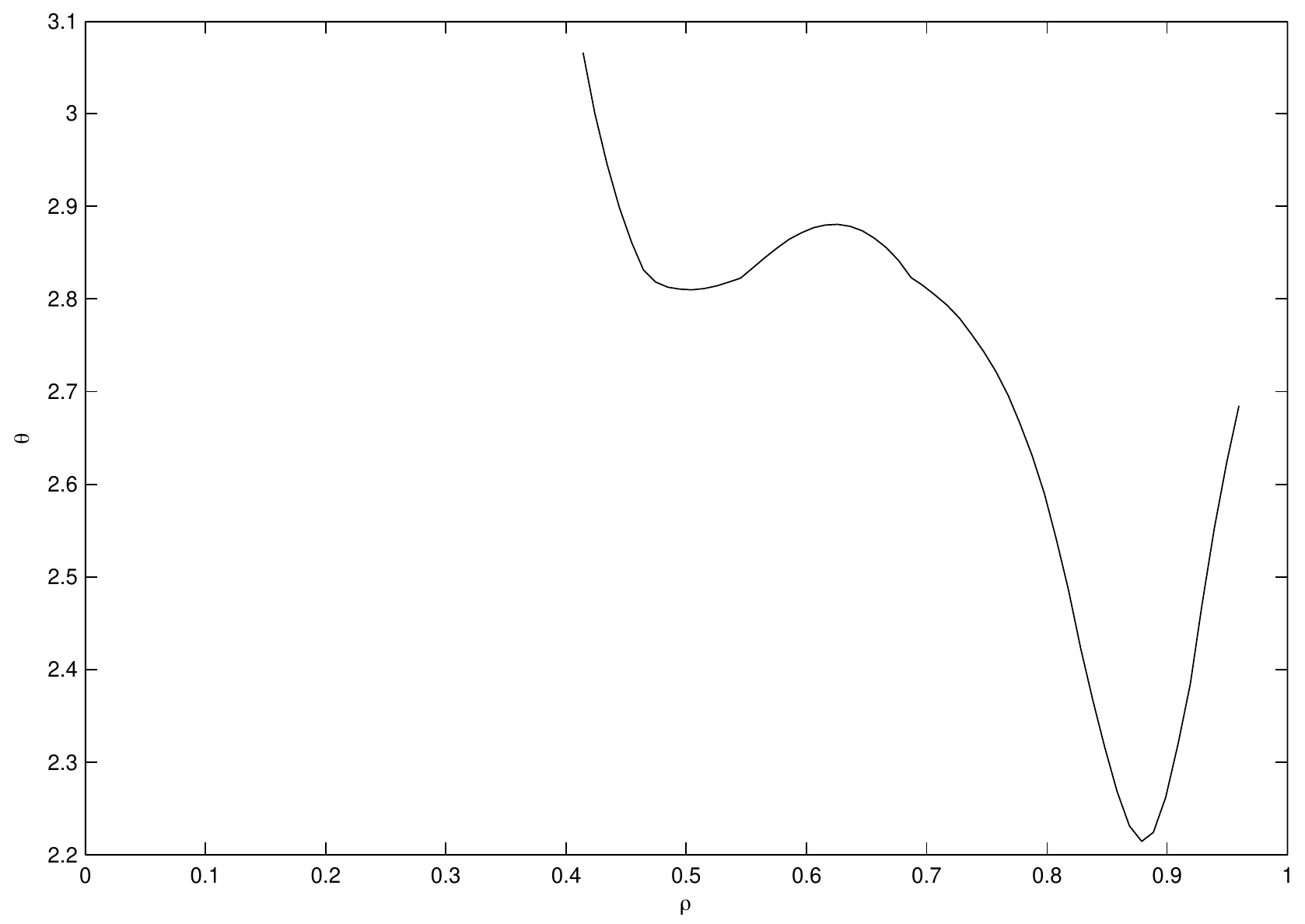}
}
\caption{Locus of the points $(r,\theta)$ where $X_i^{1}=0$. As an example of calculation, we choose the point $(\rho_{i0},\theta_{i0})=(0.5079, 2.8137)$ of this curve.}
\label{FDDi}
\end{center}
  \end{minipage}
\hfill
\end{figure*}

\begin{equation}  \label{q4}
({\hat P}^e_{\phi0},{\hat\lambda}^e_{0})=(0.1651, 0.5246)\quad; \quad ({\hat
P}^i_{\phi0},{\hat\lambda}^i_{0})=(0.5499, 0.3984)
\end{equation}
\noindent The parameters $\Delta P_\phi,\Delta\lambda_0,\Delta\lambda_1$ can
be obtained from the relations 
\begin{align}  \label{q5}
&\Bigl(\frac{\partial X^{1}_e}{\partial \rho}{\Bigg\arrowvert}%
_{\rho_{e0},\theta_{e0}},\frac{\partial X^{1}_e}{\partial \theta}\!\!{%
\Bigg\arrowvert}_{\rho_{e0},\theta_{e0}}\Bigr)=(0.0039,1.5957\times 10^{-4})
\notag \\
&\Bigl(\frac{\partial X^{2}_e}{\partial \rho}{\Bigg\arrowvert}%
_{\rho_{e0},\theta_{e0}},\frac{\partial X^{2}_e}{\partial \theta}\!\!{%
\Bigg\arrowvert}_{\rho_{e0},\theta_{e0}}\Bigr)=(-0.0028,-7.3501\times
10^{-5}) \\
&\Bigl(\frac{\partial X^{1}_i}{\partial\rho}{\Bigg\arrowvert}%
_{\rho_{i0},\theta_{i0}},\frac{\partial X^{1}_i}{\partial \theta}\!\!{%
\Bigg\arrowvert}_{\rho_{i0},\theta_{i0}}\Bigr)=(6.6596\times 10^{-4},{%
\mathcal{O}}(\epsilon^2))  \notag
\end{align}
\noindent By the chain rule for derivatives, we find 
\begin{align}  \label{q6}
&\xi_1\simeq -3.9304\times 10^{-3}\Delta {\tilde P}^e_\phi \\
&\xi_2\simeq 3.3571\times 10^{-3}\Delta {\tilde\lambda}^e  \notag
\end{align}
\noindent From which we obtain 
\begin{align}  \label{q7}
&\Delta {\hat P}_\phi^e= 182.278\qquad;\quad \Delta{\hat\lambda}%
^e_0\rightarrow\infty\quad ;\quad \Delta{\hat\lambda}^e_1=581.268 \\
&(\Delta {\hat P}_\phi^i)^{-2}\sim{\mathcal{O}}(\epsilon^2)\quad;\quad \Delta%
{\hat\lambda}^i_0\rightarrow\infty\quad ;\quad \Delta{\hat\lambda}%
^i_1=286.236  \notag
\end{align}
\noindent Hence, the electron and ion density distribution functions finally
read 
\begin{align}  \label{q8}
&{\mathcal{F}}^{eR}\propto\exp\Bigl[-\Bigl(\frac{{\hat w}-0.3155}{0.2385}%
\Bigr)^2-\Bigl(\frac{{\hat P}-0.1651}{182.278}\Bigr)^2-\Bigl(\frac{{%
\hat\lambda}-0.5246}{581.268}\Bigr)^2\ \Bigr]\mid\!{\mathcal{J}}\!\mid \\
&{\mathcal{F}}^{iR}\propto\exp\Bigl[-\Bigl(\frac{10^{3}\times{\hat w} -0.1983%
}{0.1057}\Bigr)^2-\Bigl(\frac{{\hat\lambda}- 0.3984}{286.236}\Bigr)^2\ \Bigr]%
\mid\!{\mathcal{J}}\!\mid  \notag
\end{align}
\noindent where the dimensionless variable ${\hat w}\equiv w/v_{the}^2$,
with $v_{the}$ computed at the center of the tokamak, has been introduced.

\noindent By summarizing, the reference density distribution function ${\mathcal F}^R$, given by Eq.~(\ref{ddf7}), identifies with the reference DDF estimated by the neoclassical theory for collisional tokamak-plasmas in the Onsager region when the free parameters in Eq.~(\ref{ddf7}) take the values given by Eqs.~(\ref{ex11}). In coordinates $\hat w$, ${\hat P}_\phi$ and $\hat\lambda$ (and $\psi$), the expressions of the reference DDFs are given by Eq.~(\ref{q8}). Notice that in this case $c_2^{(1)}=0$ ($\Delta{\hat\lambda}^\alpha_0\rightarrow\infty$) and the presence of the parameter $c_2^{(0)}$ (or of the parameter $\Delta{\hat\lambda}^\alpha_1$) is crucial.
\vskip0.2truecm
\section{Conclusions}\label{cs} 
\vskip 0.2truecm 
Using statistical thermodynamics approach we
have derived the general expression of the density distribution
function ${\mathcal{F}}^R$ for the case of a thermodynamic system out of equilibrium, subject to
three thermodynamic forces. The local equilibrium is fixed by imposing the following conditions :
\begin{description}
\item{{\bf i)} The minimum entropy production condition on the two Prigogine$'$s fluctuations $\alpha_1$ and $\alpha_2$};
\item{{\bf ii)} The maximum entropy principle on the variable $w$, for $\alpha_1=\alpha_2=0$}
\end{description}
\begin{description}
\item {{\bf iii)} The scale invariance of the restrictions used in the maximization of the entropy};
\item {{\bf iv)} A new mathematical ansatz, used in selecting a minimal number of restrictions and implicitly free parameters}.
\end{description}

\noindent {\it From this ansatz results a singularity of the DDF that has immediate physical interpretation in terms of the intermittency in turbulent plasmas}. 

\noindent The derived DDF, ${\mathcal{F}}^R$, is more general
than that currently used for fitting the numerical steady-state solution
obtained by simulating ICRH plasmas and for describing various scenarios of
tokamak plasmas.  The adopted procedure can be generalized for systems
subject to an arbitrary number of thermodynamic forces. By kinetic theory,
we have linked, and then fixed, the seven free parameters entering in ${
\mathcal F}^R$ with the external energy sources and the (internal) entropy
production source strength. To be more concrete, we have analyzed the case
of, fully ionized, magnetically confined plasmas.

\noindent \noindent This work gives several perspectives. Through the thermodynamical field theory (TFT) \cite{sonninoPRE} it is possible to estimate the DDF when the nonlinear contributions cannot be neglected \cite{sonnino2}. The next task should be to establish the relation between the reference DDF herein derived with the one found by the TFT. The solution of this difficult problem will contribute to provide a link between a microscopic description and a macroscopic approach (TFT).  Another problem to be solved is the possibility to improve the numerical fit by adding new free parameters according to the principles exposed in this work.
\vskip0.2truecm
\section{Acknowledgments}

One of us (G. Sonnino),
is very grateful to Prof. M.Malek Mansour, of the Universit{\'e} Libre de
Bruxelles, for his scientific suggestions and for his help in the
development of this work. 
\appendix 
\vskip0.2truecm
\section{A Source Model for Tokamak-plasmas heated by (ICRH)}\label{icrh}
\vskip0.2truecm
 Let us now re-consider the case mentioned in Section (\ref{parameter}): a low concentration of ions ${\ \!{}^3\!He}$ colliding with a thermal background plasma, composed by Deuterium and electrons and heated by ICRH. In the velocity space, the long term evolution of the distribution function for the high frequency heated ions, is governed by Eq.~(\ref{cm1}) \cite{brambilla}, (see the footnote \footnote{Here, we shall limit ourselves to analyze the so-called {\it simplified, quasi-linear, Fokker Planck equation} (QLFPE) where it is taken into account only the contribution due to the perpendicular component of the electric field, $E^+_{\perp}$, which is concordant to the direction of rotation of the minority \cite{brambilla}. Eq.~(\ref{cm4}) neglects then the contributions due to the perpendicular component of the electric field, $E^{-}_{\perp}$, which is discordant to the direction of rotation of the minority, and the one due to the parallel component of the electric field, $E_{\parallel}$.\label{approx}})
\begin{equation}\label{cm4}
\frac{\partial{\mathcal F}^m({\bf y},t)}{\partial t}=-\nabla\cdot \sum_{\alpha=e,i}{\bar{\bf S}}_c^{m\alpha}({\mathcal F}^m)-\nabla\cdot {\bar{\bf S}}_W^{m}({\mathcal F}^m)
\end{equation}
\noindent ${\bar{\bf S}}_c^{m\alpha}$ and the simplified quasi-linear term, ${\bar{\bf S}}_w^{m}$, can be written as \cite{brambilla}, \cite{cardinali}
\begin{align}\label{new1}
&{\bar{\bf S}}_c^{m\alpha}({\mathcal F}^m)
=-[\nabla\cdot ({\bar{\bar{\bf D}}}_c^{m\alpha(2)}{\mathcal F}^m)]^T+{\bar{\bf D}}_c^{m\alpha(1)}{\mathcal F}^m\\
&{\bar{\bf S}}_W^{m}({\mathcal F}^m)
=-[\nabla\cdot ({\bar{\bar{\bf D}}}_W^{m(2)}{\mathcal F}^m)]^T+{\bar{\bf D}}_W^{m(1)}{\mathcal F}^m\nonumber
\end{align}
\noindent with $T$ denoting the {\it transpose operation}. We introduce the following dimensionless coordinates, ${\hat w}$ and $\hat\lambda$
\begin{align}\label{neweq1}
&{\hat w}=\frac{w}{v_{the}^2}\qquad\quad {\rm where}\quad v_{th\alpha}=\sqrt{\frac{2T_\alpha}{m_\alpha}}\qquad \alpha=(e,i)\\
&\hat\lambda=B_0\lambda\nonumber
\end{align}
\noindent where $w$ is the kinetic energy per unit mass and $\lambda$ defined in Eqs~(\ref{ddf1})]. In these coordinates, the matrices ${\bar{\bar{\bf D}}}_c^{m\alpha(2)}$, ${\bar{\bar{\bf D}}}_W^{m(2)}$, ${\bar{\bf D}}_c^{m\alpha(1)}$ and ${\bar{\bf D}}_W^{m(1)}$ can be cast into the form
\begin{align}\label{new2}
&{\bar{\bar{\bf D}}}_c^{me(2)}=
\begin{pmatrix}
D_{cww}^{me(2)} & 0\\
0 & D_{c\lambda\lambda}^{me(2)}
\end{pmatrix}\qquad ; \qquad {\bar{\bar{\bf D}}}_c^{mi(2)}=
\begin{pmatrix}
D_{cww}^{mi(2)} & 0\\
0 & D_{c\lambda\lambda}^{mi(2)}
\end{pmatrix}\\
&{\bar{\bf D}}_c^{me(1)}=
\begin{pmatrix}
D_{cw}^{me(1)}+\frac{2}{v_{the}\sqrt{\hat{w}}}D_{cww}^{me(2)}+\frac{2\sqrt{\hat{w}}}{v_{the}}\frac{dD_{cww}^{me(2)}}{d\hat{w}}\\ 
\frac{1}{v_{the}}\sqrt{\frac{1-2\mid{\hat B}\mid{\hat\lambda}}{2\mid{\hat B}\mid{\hat\lambda} \hat{w}}}\ D_{c\lambda\lambda}^{me(2)}
\end{pmatrix}
\nonumber\\
&{\bar{\bf D}}_c^{mi(1)}=
\begin{pmatrix}
D_{cw}^{mi(1)}+\frac{2}{v_{the}\sqrt{\hat{w}}}D_{cww}^{mi(2)}+\frac{2\sqrt{\hat{w}}}{v_{the}}\frac{dD_{cww}^{mi(2)}}{d\hat{w}}\\
\frac{1}{v_{the}}\sqrt{\frac{1-2\mid{\hat B}\mid{\hat\lambda}}{2\mid{\hat B}\mid{\hat\lambda} \hat{w}}}\ D_{c\lambda\lambda}^{mi(2)}
\end{pmatrix}\nonumber
\end{align}
\noindent with $\mid{\hat B}\mid\equiv\mid B\mid/B_0$, and
\begin{align}\label{new3}
&{\bar{\bar{\bf D}}}_W^{m(2)}=
\begin{pmatrix}
D_{Www}^{m(2)} & D_{Ww\lambda}^{m(2)}\\
D_{W\lambda w}^{m(2)} & D_{W\lambda\lambda}^{m(2)}\\
\end{pmatrix}\\
&{\bar{\bf D}}_W^{m(1)}=\begin{pmatrix}
 \frac{2}{v_{the}\sqrt{{\hat w}}}D_{Www}^{m(2)}+\frac{2\sqrt{\hat{w}}}{v_{the}}\frac{dD_{Www}^{m(2)}}{d\hat{w}}+\frac{1}{v_{the}}\sqrt{\frac{1-2\mid B\mid{\hat\lambda}}{2\mid{\hat B}\mid{\hat\lambda} \hat{w}}}\ D_{Ww\lambda}^{m(2)}+\frac{1}{v_{the}}\sqrt{\frac{2{\hat\lambda}(1-2\mid{\hat B}\mid{\hat\lambda})}{\mid{\hat B}\mid \hat{w}}}\frac{dD_{Ww\lambda}^{m(2)}}{d{\hat\lambda}}\\
\frac{2}{v_{the}\sqrt{{\hat w}}}D_{W\lambda w}^{m(2)}+\frac{2\sqrt{\hat{w}}}{v_{the}}\frac{dD_{W\lambda w}^{m(2)}}{d\hat{w}}+\frac{1}{v_{the}}\sqrt{\frac{1-2\mid{\hat B}\mid{\hat\lambda}}{2\mid{\hat B}\mid{\hat\lambda} \hat{w}}}\ D_{W\lambda\lambda}^{m(2)}+\frac{1}{v_{the}}\sqrt{\frac{2{\hat\lambda}(1-2\mid{\hat B}\mid{\hat\lambda})}{\mid{\hat B}\mid \hat{w}}}\frac{dD_{W\lambda\lambda}^{m(2)}}{d{\hat\lambda}}
\nonumber
\end{pmatrix}
\end{align}
\noindent Notice that in Eqs~(\ref{new2}) and (\ref{new3}), extra terms appear in the components of the drift vectors. However, the nature of these additional terms is {\it purely geometrical} and they come from the conversion of the balance equation, Eq.~(\ref{cm4}), into the Fokker-Planck equation re-written in the standard way. The expressions of the collisional electron and ion coefficients are \cite{brambilla}, \cite{cardinali}
\begin{align}\label{new4}
&D_{cww}^{me(2)}=\frac{\Gamma^{me}}{2v_{the}\sqrt{\hat w}}\Bigl(\frac{{\rm erf}({\sqrt{\hat w}})-{\sqrt{\hat w}}\ {\rm erf}'({\sqrt{\hat w}})}{{\hat w}}\Bigr)\\
&D_{c\lambda\lambda}^{me(2)}=\frac{\Gamma^{me}}{4v_{the}\sqrt{\hat w}}\Bigl(\Bigl(2-\frac{1}{\hat w}\Bigr){\rm erf}({\sqrt{\hat w}})+\frac{{\rm erf}'({\hat w})}{\sqrt{\hat w}}\Bigr)\nonumber\\
&D_{cw}^{me(1)}= -\frac{\Gamma^{me}}{v^2_{the}}\frac{m_i}{m_e}\frac{[{\rm erf}(\sqrt{\hat w})-\sqrt{\hat w}\ {\rm erf}'(\sqrt {\hat w})]}{\hat w}\qquad\qquad {\rm where}\nonumber\\
& \Gamma^{m\alpha}=\frac{4\sqrt{2\pi}}{3}\frac{ n_\alpha e^4Z_m^2Z_\alpha^2\ln\Lambda_{m\alpha}}{m_m^2}\ \ ; \ \ \Lambda_{m\alpha}=\frac{3(T_m+T_\alpha)\lambda_{Dm\alpha}}{2Z_mZ_\alpha e^2}
\nonumber\\
&\lambda_{Dm\alpha}=\Bigl(\frac{T_m T_\alpha(Z_m+Z_\alpha)}{4\pi Z_mZ_\alpha e^2(n_mT_m+n_\alpha T_\alpha)}\Bigr)^{1/2}
\nonumber
\end{align}
\noindent and
\begin{align}\label{new5}
&D_{cww}^{mi(2)}=\frac{\Gamma^{mi}}{2v_{the}\sqrt{\hat w}}\Bigl(\frac{{\rm erf}({\sqrt{\kappa\hat w}})-{\sqrt{\kappa\hat w}}\ {\rm erf}'({\sqrt{\kappa\hat w}})}{\kappa{\hat w}}\Bigr)\\
&D_{c\lambda\lambda}^{mi(2)}=\frac{\Gamma^{mi}}{4v_{the}\sqrt{\hat w}}\Bigl(\Bigl(2-\frac{1}{\kappa\hat w}\Bigr){\rm erf}({\sqrt{\kappa\hat w}})+\frac{{\rm erf}'({\sqrt{\kappa\hat w}})}{\sqrt{\kappa\hat w}}\Bigr)\nonumber\\
&D_{cw}^{mi(1)}= -\frac{\Gamma^{mi}}{v^2_{the}}\frac{({\rm erf}(\sqrt{\kappa\hat w})-\sqrt{\kappa\hat w}\ {\rm erf}'(\sqrt {\kappa\hat w}))}{\hat w}\quad ;\quad {\rm where}\qquad \kappa\equiv\frac{v_{the}^2}{v_{thi}^2}\nonumber
\end{align}
\noindent respectively. Notice that, the quantities in Eqs~(\ref{new4}), are expressed in the {\it CGS Gaussian units}. ${\rm erf}(x)$ and ${\rm erf}'(x)$ denote the {\it error function} and the {\it derivative} of the error function with respect to its argument, respectively. The expressions of the diffusion coefficients for the resonant wave particle interactions are \cite{brambilla}, \cite{cardinali}
\begin{align}\label{new6}
&D_{Www}^{m(2)}=2\mid{\hat B}\mid{\hat\lambda}\ D_{\perp\perp}^m\\
&D_{Ww\lambda}^{m(2)}=D_{W\lambda w}^{m(2)}=\sqrt{(2\mid{\hat B}\mid{\hat\lambda})(1-2\mid{\hat B}\mid{\hat\lambda})}\ D_{\perp\perp}^m\nonumber\\
&D_{W\lambda\lambda}^{m(2)}=(1-2\mid{\hat B}\mid{\hat\lambda})\  D_{\perp\perp}^m\nonumber\\
&D_{\perp\perp}^m=D_0^mJ^2_p(\xi_m\sqrt{2\mid{\hat B}\mid{\hat\lambda}{\hat w}})\qquad\qquad\qquad\qquad{\rm with}\nonumber\\
&\xi_m=\frac{v_{the}\kappa_\perp}{\Omega_{cm}}\quad ;\quad \Omega_{cm}=\frac{Z_me B_0}{ m_mc}\quad;\quad 
P_{abs}^{lin}\simeq10\div 80\ Watt/cm^3\nonumber\\
&D_0^m=\frac{P_{abs}^{lin}}{4m_mn_m\int_0^\infty x^3J_p^2(\xi_mx)\exp{(-x^2)}\ dx}\nonumber
\end{align}
\noindent Here, $J_p(x)$ indicates the Bessel functions of the first kind. $e$ and $c$ are the absolute value of the charge of the electron and the speed of light, respectively. From Eqs~(\ref{new2})-(\ref{new6}) we obtain the expressions of the {\it total matrix diffusion coefficients} and the {\it total drift vector coefficients}
\begin{align}\label{new7}
&{\bar{\bar{\bf D}}}^{m(2)}={\bar{\bar{\bf D}}}_c^{me(2)}+{\bar{\bar{\bf D}}}_c^{mi(2)}+{\bar{\bar{\bf D}}}_W^{m(2)}=
\begin{pmatrix}
D_{ww}^{m(2)} & D_{w\lambda}^{m(2)}\\
D_{w\lambda}^{m(2)} & D_{\lambda\lambda}^{m(2)}\\
\end{pmatrix}\\
&{\bar{\bf D}}^{m(1)}={\bar{\bf D}}_c^{me(1)}+{\bar{\bf D}}_c^{mi(1)}+{\bar{\bf D}}_W^{m(1)}=
\begin{pmatrix}
D_{w}^{m(1)}\\ 
D_{\lambda}^{m(1)}
\end{pmatrix}
\nonumber
\end{align}
\noindent In the zero orbit width limit, the steady state DDF of the Fokker-Planck equation, with the diffusion and drift coefficients given by Eqs~(\ref{new7}), may be fitted by the profile
\begin{equation}\label{new8}
{\mathcal F}^m\propto \Bigl(\frac{w}{\Theta}\Bigr)^{-3/4}\exp\Bigl[-\Bigl(\frac{P-P_\phi}{\Delta P_\phi}\Bigr)^2\Bigr]
\exp\Bigl(-\frac{w}{\Theta}\Bigr)\delta(\lambda-\lambda_0)\mid\!{\mathcal J}\!\mid
\end{equation}
\noindent with $\delta(\lambda-\lambda_0)$ indicating the Dirac function. Figs~(\ref{FDD_D2ww}), (\ref{FDD_D2wl}) and (\ref{FDD_D2ll}), report on the graphics of the dimensionless elements of the total diffusion matrix, ${\hat D}^{m(2)}_{ww}$, ${\hat D}^{m(2)}_{w\lambda}={\hat D}^{m(2)}_{\lambda w}$ and ${\hat D}^{m(2)}_{\lambda\lambda}$ [see the first equation in Eqs~(\ref{new7})], versus $\hat w$ and $\hat\lambda$, respectively. The dimensionless diffusion coefficients are defined as ${\hat D}^{m(2)}_{ij}\equiv D^{m(2)}_{ij}v_{the}/\Gamma^{me}$.
\begin{figure}\resizebox{0.55\textwidth}{!}{%
\includegraphics{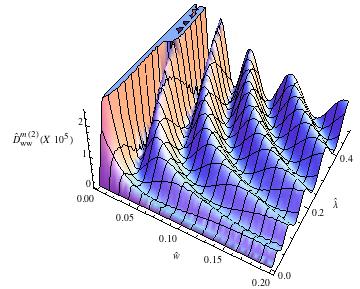}}
\caption{ \label{FDD_D2ww} Dimensionless diffusion component ${\hat D}^{m(2)}_{ww}$ versus the normalized, dimensionless variables ${\hat w}$ and $\hat\lambda$. The dimensionless diffusion coefficients are defined as ${\hat D}^{m(2)}_{ij}\equiv D^{m(2)}_{ij}v_{the}/\Gamma^{me}$.}
\end{figure}
\begin{figure}\resizebox{0.55\textwidth}{!}{%
\includegraphics{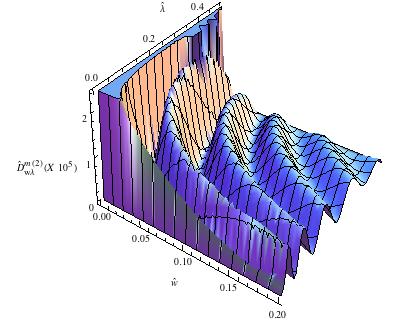}}
\caption{ \label{FDD_D2wl} Dimensionless diffusion component ${\hat D}^{m(2)}_{w\lambda}$ versus ${\hat w}$ and $\hat\lambda$. The matrix of the diffusion coefficients is symmetric: ${\hat D}^{m(2)}_{w\lambda}={\hat D}^{m(2)}_{\lambda w}$.}
\end{figure}
\begin{figure}\resizebox{0.55\textwidth}{!}{%
\includegraphics{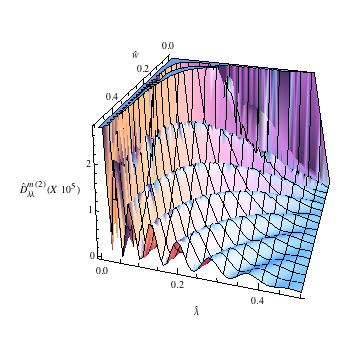}}
\caption{ \label{FDD_D2ll} Dimensionless diffusion component ${\hat D}^{m(2)}_{\lambda\lambda}$ versus ${\hat w}$ and $\hat\lambda$.}
\end{figure}

\noindent Figs~(\ref{FDD_D1w})-(\ref{FDD_D1l}) report on the dimensionless components of the total drift coefficients [see the second equation in Eqs~(\ref{new7})], defined as ${\hat D}^{m(1)}_{i}\equiv D^{m(1)}_{i}v^2_{the}/\Gamma^{me}$, against $\hat w$ and $\hat\lambda$. 
\begin{figure}\resizebox{0.55\textwidth}{!}{%
\includegraphics{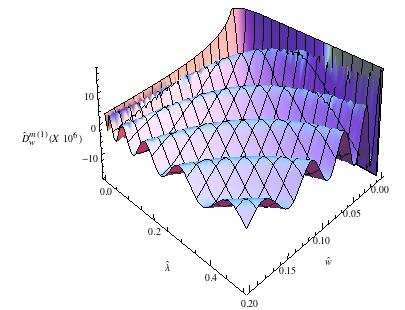}}
\caption{ \label{FDD_D1w} Plot of the first dimensionless component of the total drift vector, ${\hat D}^{m(1)}_{w}$, versus ${\hat w}$ and $\hat\lambda$.}
\end{figure}
\begin{figure}\resizebox{0.55\textwidth}{!}{%
\includegraphics{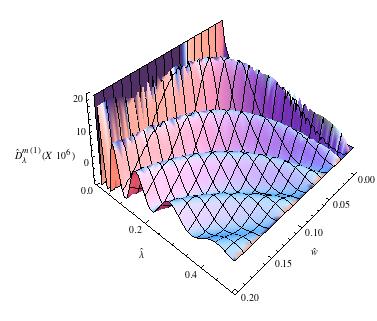}}
\caption{ \label{FDD_D1l} Plot of the second dimensionless component of the total drift vector, ${\hat D}^{m(1)}_{\lambda}$, versus ${\hat w}$ and $\hat\lambda$.}
\end{figure}

\noindent These pictures have been obtained by considering only the fundamental cyclotron heating, $p=0$, (minority heating) and by setting $\kappa_\perp\simeq \rho_{Lm}^{-1}$ (with $\rho_{Lm}$ denoting the {\it Larmor gyroradius}). Moreover, as an example of calculation, we chose the following values of the tokamak parameters: $B_0=3.45\ Tesla$, $R_0=2,96\ m$ and $a=1,25\ m$. Moreover, $P_{abs}^{lin}=50\ Watt/cm^3$ and $n_m=2.5\%\ n_i$, respectively and, as an example of calculation, the fields have been estimated at the center of the tokamak. Notice that in our case, the charge number and the ion mass are $Z_\alpha=1$ and $m_i=m_{D}=2m_{H}$ for the background, and $Z_m=Z_{\ \!{}^3\!He}=2$ and $m_m=m_{\ \!{}^3\!He}\simeq 3 m_H$ (with $m_H$ indicating the mass of proton) for the minority, respectively. 
\vskip0.2truecm
\section{The Physical Justification of the Gamma Distribution Function}\label{just}
\vskip0.2truecm
In order to justify the $w$ dependence of $\mathcal{P}(w)$ we will use the principle of maximal entropy with suitable
chosen restrictions obtained from scale invariance and a new mathematical ansatz. We denote by $S[\rho (.)]$ the entropy functional of a
given probability density function $\rho (w)$ : 
\begin{equation}
S[\rho (.)]=-\int_{0}^{\infty }\rho (w)\log (\rho (w))dw  \label{SG1}
\end{equation}%
\noindent The maximum entropy principle will be applied by imposing a set of
restrictions: 
\begin{align}
\ \int_{0}^{\infty }\rho (w)dw& =1  \label{SG2} \\
\ \int_{0}^{\infty }w\rho (w)dw& =\mathrm{E}(w)=\mu _{1}  \label{SG3} \\
\ \int_{0}^{\infty }\log (w)\rho (w)dw& =\mathrm{E}(\log (w))=\nu
\label{SG4}
\end{align}%
\noindent The naturalness of the restriction given by Eq.~(\ref{SG3}) is
clear: we observe that $\mu _{1}=T/m$ where $\ T$ is the temperature.
Concerning the naturalness of the choice of restriction (\ref{SG4}), we
remark that all of the previous restrictions belongs to a class that is
invariant under scaling. More generally, for a restriction of the form $%
\mathrm{E}(w^{\alpha })=\mu _{\alpha }$, respectively for $\mathrm{E}(\log
(w))=\nu $, the effect scaling $w=kw^{\prime }$ is $\mathrm{E}(w^{\prime
\alpha })=\mu _{\alpha }^{\prime }=\mu _{\alpha }k^{-\alpha }$, respectively
for $\mathrm{E}(\log (w^{\prime }))=\nu ^{\prime }=\nu -\log (k)$. The
problem why the PDF obtained by the one of the possible simplest choice from
the more general set of restrictions 
\begin{align}
& \mathrm{E}(w^{\alpha _{k}})=\mu _{k}~;\qquad \mathrm{with}\qquad \mu
_{0}\equiv 1;~\alpha _{0}=0;\alpha _{1}=1;k=0,1,2,\ldots ,  \label{SG5} \\
& \mathrm{E}(\log (w))=\nu  \label{SG6}
\end{align}%
\noindent deserves further study. A partial answer is given by simplicity
and extremality reasons. Observe first that the restriction (\ref{SG6}) can
be seen as a limiting case of (\ref{SG5}). Indeed, suppose that we have for
some fixed $k$ : $\alpha _{k}=\varepsilon \ll 1$ 
\begin{equation}
\mathrm{E}(w^{\varepsilon })=\mu _{k}~  \label{SG7}
\end{equation}%
\noindent From (\ref{SG2}) and (\ref{SG7}) results 
\begin{equation}
\mathrm{E}\left( \frac{w^{\varepsilon }-w^{0}}{\varepsilon }\right) =\frac{%
\mu _{k}-1}{\varepsilon }  \label{SG8}
\end{equation}%
\noindent When the support of the PDF $\rho (w)$ is concentrated mainly on
the domain when $|\log (w)|$ is not too large, then we can approximate: $%
\left( w^{\varepsilon }-w^{0}\right) /\varepsilon \cong \log (w)$ , so the
Eq.(\ref{SG8}) is reduced to Eq.(\ref{SG4}). So, the restriction of the type
(\ref{SG4}) can be seen as the an extreme case. It is easily checked that
applying the maximal entropy principle from \ref{SG6} results $\rho (w)\underset{w\rightarrow 0}{\asymp }{\rm const}\ w^{\gamma -1}$ for some $\gamma
>0$. Notice that when $\gamma <1$, the singularity of our DDF is related to the \emph{intermittency} shown by real physical DDFs \cite{Aumaitre}, with time and ensemble average are provided by our DDF. On the other hand, it is easily checked that in the case of restrictions, %
\ref{SG5} after applying the maximal entropy principle, we obtain $\rho (w)%
\underset{w\rightarrow \infty }{\asymp }c_{1}\exp (-c_{2}w^{m})$ for
some $c_{1},c_{2}>0$, where $m=\max \{\alpha _{1},...,$ $\alpha _{n}\}$. But $m>1$ give a much faster decay compared to Maxwell distribution, so we
obtain $m<1$. {\it Consequently, we use the minimal ansatz given by} Eqs~(\ref{SG2})-(\ref{SG4}), {\it for the selection of the restrictions}. From
this reason of extremality of the restriction Eq.(\ref{SG4}) and motivated
by the minimality assumptions, we will explore the consequences of the
minimal and extremal model given by Eqs~(\ref{SG2})-(\ref{SG4}). In order to
find the distribution function $\rho (w)$ that maximizes the entropy (\ref%
{SG1}), with restrictions (\ref{SG2})-(\ref{SG4}) we use the Lagrange
multiplier method. The result is the gamma distribution function, given by \cite{sonninopreb}
\begin{equation}
\rho (w)=\frac{\Theta }{\Gamma (\gamma )}\left( w/\Theta \right) ^{\gamma
-1}\exp \left( -w/\Theta \right)   \label{SG9}
\end{equation}%
\noindent where $\Gamma (\gamma )$ is the Euler gamma function. The relation
between the parameters $\gamma $, $\Theta $ and $\mu _{1}=T/m$, $\nu $ is
given by 
\begin{align}
\ \ \mathrm{E}(w)& =T/m=\gamma \Theta   \label{SG10} \\
\ \mathrm{E}(\log (w))& =\nu =\Psi (\gamma )+\log (\Theta )  \label{SG11}
\end{align}
\noindent where $\Psi (\gamma )$ is the digamma function. Due to the very
special mathematical peculiarity in the ansatz Eqs~(\ref{SG2})-(\ref{SG4}), the resulting PDF is
expected to have also some special properties. Indeed, the gamma
distribution is \emph{infinitely divisible and stable }: if the PDF of
independent random variables $X_{1},\ldots ,X_{n}$ is a gamma distribution
with the same scale parameter and shape parameters $\gamma _{1},\ldots
\gamma _{n}$, then PDF of the random variable $\sum_{k=1}^{n}X_{k}$ is again
a gamma distribution with the same scale parameter and with shape parameter $%
\sum_{k=1}^{n}\gamma _{k}$. We are convinced that the physical interpretation of the infinite divisibility property is a challenging problem.





\begin{thebibliography}{99}

\bibitem{balescu0} R. Balescu, 1977 \textit{Statistical Dynamics. Matter out
of Equilibrium}, Imperial College Press, Printed in Singapore by Uto-print.

\bibitem{sonninoPRE1} G. Sonnino, G. Steinbrecher, A. Cardinali, A. Sonnino and M. Tlidi \textit{Phys.Rev. E}, \textbf{87}, 014104, (2013).

\bibitem{prigogine} I. Prigogine (1947), \textit{Etude Thermodynamique des
Ph{\`e}nom{\`e}nes Irr{\'e}versibles}, Th{\`e}se d'Aggr{\'e}gation de
l'Einseignement Sup{\'e}rieur de l'Universit{\'e} Libre de Bruxelles
(U.L.B.).

\bibitem{prigogine1} I. Prigogine, 1954 \textit{Thermodynamics of
Irreversible processes}, (John Wiley \& Sons).

\bibitem{degroot} S.R. De Groot and P. Mazur, 1984 \textit{Non-Equilibrium
Thermodynamics}, Dover Publications, Inc., New York.

\bibitem{sonninoPRE} G. Sonnino \textit{Phys.Rev. E}, \textbf{79}, 051126,
(2009).

\bibitem{bilato} M. Brambilla and R. Bilato, {\it Nucl. Fusion}, {\bf 49}, 085004 (2009).

\bibitem{cardinali1} A. Cardinali \textit{et al.}, \textit{23rd IAEA Fusion
Energy Conference}, (2010).

\bibitem{pizzuto} A. Pizzuto \textit{et al.}, \textit{Nucl. Fusion}, \textbf{%
50}, 95005, (2010).

\bibitem{zonca} F. Zonca and L. Chen, \textit{Phys. Plasmas}, \textbf{7},
4600, (2000).

\bibitem{balescu2} R. Balescu, 1988 \textit{Transport Processes in Plasmas.
Vol 2. Neoclassical Transport}, Elsevier Science Publishers B.V., Amsterdam,
North-Holland.

\bibitem{balescu1} R. Balescu, 1988 \textit{Transport Processes in Plasmas.
Vol 1. Classical Transport}, Elsevier Science Publishers B.V., Amsterdam,
North-Holland.

\bibitem{park} Y. Park, Sung, K. Bera, Anil, \textit{Journal of Econometrics}
(Elsevier), 219 (2009). Retrieved (2011).

\bibitem{papoulis} A. Papoulis and S.U. Pillai, 2002 \textit{Probability,
Random Variables and Stochastic Processes}, Mac Graw Hill, Fourth edition.

\bibitem{gardiner} C. W. Gardiner, 2004 \textit{Handbook of Stochastic
Methods: for Physics, Chemistry and the Natural Sciences}, Third edition
Springer Verlag, Berlin, Heidelberg, New York.

\bibitem{sonninopreb} G. Sonnino, G. Steinbrecher, A. Cardinali, A. Sonnino and M. Tlidi \textit{Phys.Rev. E}, \textbf{87}, 014104, (2013).

\bibitem{sonninocpp} G. Sonnino, {\it Contributions To Plasma Physics} (CPP), {\bf 51}, 798 (2011).

\bibitem{brambilla} M. Brambilla, {\it Nuclear Fusion}, {\bf 34}, No.8, 1121 (1994).

\bibitem{cardinali} A. Cardinali, S. Briguglio and G. Calabr{\`o}, {\it et Al.}, {\it Nuclear Fusion}, {\bf 49}, 095020 (2009).

\bibitem{sonnino} G. Sonnino and P. Peeters \textit{Physics of Plasmas} 
\textbf{15}, 062309/1-062309/23 (2008).

\bibitem{sonnino2} G. Sonnino, \textit{The European Physical Journal D}
(EPJD), \textbf{62}, Issue 1, 81, (2011).

\bibitem{Aumaitre} S. Aumaitre, F. Petrelis, K. Mallick, \textit{Phys. Rev.Lett.} \textbf{95}, Issue 6, 064101, (2005).

\end{thebibliography}
\end{document}